\documentclass{jfm}
\usepackage{graphicx}
\usepackage{epstopdf, epsfig}

\usepackage{color}

\def\tb{\textcolor{black}}

\title{\tb{Particle trapping in} vortex crystals} 

\author{Jean-R\'egis Angilella\aff{1}
\corresp{\email{Jean-Regis.Angilella@unicaen.fr}}
\and S. Ravichandran\aff{2}}

\affiliation{\aff{1}Université de Caen Normandie and CNRS UMR 7190, Institut Jean Le Rond d'Alembert, Paris, France
\aff{2}Centre for Climate Studies, Indian Institute of Technology Bombay, Powai, Mumbai 400076, India}   
  
\pubyear{}
\volume{}
\pagerange{}

\begin{document}
\maketitle
\begin{abstract}
We study the motion of inertial particles in two-dimensional inviscid and viscous vortex crystals, where vortices are placed at the tips of a regular polygon, and determine the conditions under which long-term trapping of particles can occur. In crystals with a central vortex, we find that trapping points in addition to those found in a previous analysis may exist finding, furthermore, that particle trajectories may approach a limit-cycle which can be described by means of asymptotic methods. We show that these newly discovered fixed points persist temporarily in the presence of viscosity, and study the different transitions following viscous vortex mergers. We find that for moderate Reynolds numbers, the annular merger of the vortex crystal form an annular vortex layer that can keep particles trapped within a disk centered at the origin. For larger Reynolds numbers, pairwise vortex merger leads to crystals with a smaller number of vortices, with particles clustering at the fixed points of the newly formed vortex crystal. 
\end{abstract}
  
\begin{keywords}
Inertial particles, vortical flows.
\end{keywords}

   \bibliographystyle{jfm}

\section{Introduction}
\label{secIntro}
     
 The mission of the Juno spacecraft around Jupiter in 2016 renewed the interest of the fluid physics community for vortex crystals, as it revealed spectacular vortex structures at the poles of the gaseous planet \citep{Adriani2018}. Vortex crystals are peculiar point vortex configurations that emerge in inviscid two-dimensional fluid flows, where identical vortices place themselves in regular positions, forming a kind of rigid body that rotates at a well-defined angular velocity depending on the  intensity of the vortices and on the distance between them \citep{Aref2002review}.   
  In the past decades, these structures were observed in various laboratory flows with different materials and scales: superfluid Helium \citep{Yarmchuk1979}, pure electron columns \citep{Fine1995,Schecter1999,Durkin2000PoF,Durkin2000PRL},   rotating disks floating on a liquid \citep{Grzybowski2000,Grzybowski2002}. They can emerge spontaneously from disordered vortex distributions  
  in plane flows at large Reynolds numbers \citep{Fine1995,Jin2000,Siegelman2022,Vankan2024}, as well as in forced two-dimensional turbulence \citep{Jimenez2007,Jimenez2020}.   
  These studies showed a wide variety of vortex crystals, as vortices can either place themselves at the tips of a rotating polygon (with or without an additional vortex at the center of this polygon), or form more space filling structures composed of embedded polygons. 
 
 The transport of inertial particles in these flows is a topic of interest, as particles are very sensitive to the distribution of vorticity.   Throughout the paper it will be assumed that the suspension is dilute and that the dispersed phase does not affect the fluid flow.   When particles are heavier than the fluid, they are known to flee vortex cores because of centrifugal effects.   However, when multiple vortices interact, particles can accumulate in various well-defined zones near vortices \citep{Vilela2007,Angilella2010,Ravichandran2014, Nizkaya2010,Zhao2025}.  
 The simplest case is the pair of identical point vortices, which rotate around each other until vortex merger takes place. Heavy particles released in this flow can be trapped by two attracting points rotating with the vortex pair \citep{Angilella2010,Ravichandran2014}. This is due to the fact that, in the frame rotating with the pair, anticyclonic recirculation cells are present and the Coriolis force acts towards the interior of the cell (see also \cite{Tanga1996} for a description of this phenomenon in accretion disks).  Particles with a sufficiently small Stokes number, measuring particle inertia (defined below), will then be driven towards a stable equilibrium point  located within the cell. 
 
Anticyclonic cells are also present in the flow due to crystals of vorticity composed of more than two vortices, when the flow is observed in the  frame rotating with the crystal. 
   To our knowledge, little has been done to understand the dynamics of particles in such systems. 
The question of the existence of trapping points within these cells is therefore of interest and has been discussed by \cite{Ravichandran2017}. These authors focused on the case where the crystal is composed of $N$ identical vortices placed at the tips of a polygon, i.e. regularly placed on the circumscribed circle of the polygon,
and could show numerically that attracting points exist for all $N \le 7$. They also calculated numerically the critical Stokes number below which trapping happens. The nature of trapping points (hyperbolic or degenerate)  has been investigated recently \citep{Angilella2024PoF}.  Particles with a sufficiently small Stokes number were observed to converge towards $N$ external hyperbolic trapping points located outside the circumscribed circle. 
In addition, an unexpected ``cage effect" was also observed, where particles {attracted towards the degenerate fixed point at the origin} slowly converged towards the center of the polygon.  At long times, particles therefore form a {\it mass crystal}, that is a rigid body, composed of $N+1$ points fixed in the frame of the vortex crystal. In these analyses, the flow had no vortex at the center of the polygon. This configuration has been shown to be stable for $N \le 7$ \citep{Mertz1978,Kurakin2002}. It is characterized by $N+1$ anticyclonic cells: $N$ cells with a center located outside the circumscribed circle, plus one cell around the center of the crystal.

The present work is devoted to the case where the polygon formed by the vortex crystal has a vortex located at its center. Indeed, geophysical observations, like laboratory experiments, have shown that this stable configuration is frequent \citep{Adriani2018,Durkin2000PoF}. Its stability, in the case of point vortices, has been shown theoretically \citep{Mertz1978}.
Clearly, the cage effect mentioned above and the structure of the particle crystal must be significantly affected by the presence of the central vortex. This transport phenomenon will be studied in two steps. As mentioned above, in the limit of small Stokes numbers, particles are known to possess equilibrium points in the vicinity of the center of anticyclonic recirculation cells \citep{Angilella2024PoF}. We will therefore study the number and position of such cells as a first step (section \ref{secflow}). Then, the motion of inertial particles in these cells, and the stability of their equilibrium positions, will be investigated (section \ref{secnum}). In addition, it will be shown that an attracting streamline exists inside the crystal: this one-dimensional attractor will be characterized in section \ref{secattract_LdC}. The effects of viscosity on the flow and particle dynamics will be considered in \S \ref{sec:viscous} in various configurations of interest, with or without a central vortex. 
We summarize our results and conclude in \S \ref{sec:discussion}.

\section{Anticyclonic cells and particle trapping in  vortex crystals}
\label{secflow}

\subsection{Inviscid flow structure }
 
\begin{figure}
\centerline{\includegraphics[width=0.4\textwidth]{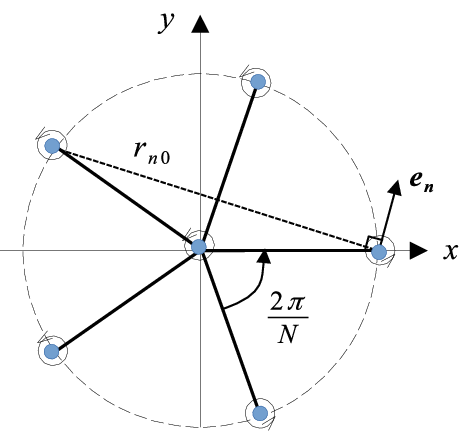}}
\caption{Sketch of  vortices placed on the tips of a polygon.}
\label{SketchCrystal}       
\end{figure}

We consider the two-dimensional flow of an inviscid and incompressible fluid, composed of  $N$ identical point vortices with strength $\Gamma > 0$, initially located at $\mathbf x_n = a (\cos(n 2\pi/N), \sin(n 2\pi/N))$, with $n=0,..,N-1$, in the Cartesian axes attached to the laboratory frame. In addition, a central vortex with strength $\Gamma_c$ is located at $\mathbf x = \mathbf 0$. The velocity induced on vortex $n=0$ by all other vortices is
\begin{equation}
\mathbf V_0 = \frac{\Gamma}{2\pi} \sum_{n\not = 0} \frac{\mathbf e_n}{r_{n0}} + \frac{\Gamma_c}{2\pi a} \mathbf e_y
\end{equation}
where $\mathbf e_n$ is the direct unit vector perpendicular to $\mathbf x_0 - \mathbf x_n$, and
$r_{n0}=|\mathbf x_0 - \mathbf x_n|=2 a \sin(n \pi/N)$, and $\mathbf e_y$ is the unit vector along the $y$ axis (figure \ref{SketchCrystal}). One can check that $\mathbf V_0 = a \Omega_0 \, \mathbf e_y$, where
\begin{equation}
\Omega_0 = (N-1)\frac{\Gamma}{4\pi a^2} + \frac{\Gamma_c}{2\pi a^2}.
\end{equation}
By symmetry, all vortices rotate at the same speed $\Omega_0$, which is the angular velocity of the crystal.

Throughout the paper, relative velocity, relative streamfunction and relative streamlines will refer to these kinematic quantities observed in the  reference frame of the vortex crystal, also referred to as "the rotating frame".
In this frame, vortices are fixed, the flow is steady and its streamfunction reads  
\begin{equation}
\psi = \frac{1}{2}\Omega_0 |\mathbf x|^2 - \frac{\Gamma}{4 \pi} \sum_n \ln(|\mathbf x - \mathbf x_n|^2)
- \frac{\Gamma_c}{4 \pi}   \ln(|\mathbf x|^2).
\label{streamfun}
\end{equation}
The first term in Eq.\ (\ref{streamfun}) is the streamfunction of a solid body rotating at speed $-\Omega_0$ with respect to the crystal, the second term corresponds to the potential flow due to the satellite vortices, and the latter is the central vortex. The flow in the rotating frame is rotational, with a uniform vorticity $-\Delta\psi = -2 \Omega_0$. In the remainder of the manuscript, {dimensional quantities} are non-dimensionalised using $a$ (the radius of the circumscribed circle of the polygon) and $\Omega_0$ (the angular velocity of the crystal with respect to the laboratory frame). Also, \tb{we introduce the vortex strength ratio} $\gamma_c = \Gamma_c/\Gamma$.

\subsection{Dynamics of particles and equilibrium points}

The non-dimensional equation of motion for isolated heavy inertial particles transported in this steady flow, in the limit of small particle Reynolds number and neglecting gravity, reads
\begin{equation}
\ddot {\mathbf X} = \frac{1}{St} (\mathbf u(\mathbf X) - \dot {\mathbf X} ) +  \mathbf X - 2 \mathbf{e}_z \times \dot {\mathbf X} 
\label{eqmvt}
\end{equation}
where $\mathbf{X}(t)$ is the position of the particle and 
$\mathbf u   = \nabla \psi  \times \mathbf{e}_z$.    Added mass and Basset-Boussinesq forces, as well as pressure gradient of the undisturbed flow (including pressure gradient due to inertial forces acting on the fluid), have been neglected in the heavy-particle limit. 
The second term on the right-hand-side of Eq.\ (\ref{eqmvt}) is the non-dimensional centrifugal force due the constant rotation of the reference frame, and the last term is the non-dimensional Coriolis force.
The Stokes number  is $St= \Omega_0 \tau_p$, where $\tau_p$ is the response time of the particles :  $\tau_p = m_p/(6 \pi \eta r_p)$, where $\eta$ is the  dynamic viscosity of the fluid, and $m_p$  and $r_p$ denote the mass and radius of the particle.  
Equilibrium positions of inertial particles are points where   drag is balanced by the centrifugal force, and satisfy
\begin{equation}
 \mathbf \Phi(\mathbf X,St) \equiv  \mathbf u(\mathbf X)   + St\,  \mathbf X  = {\mathbf 0} .
 \label{PhiEquil}
\end{equation}
The critical Stokes number below which Eq.\ (\ref{PhiEquil}) admits solutions $\mathbf X_{eq}$ could be obtained analytically in the cases $N=2$ and $N=3$
\citep{Angilella2010, Angilella2024PoF}
and numerically for larger $N$ \citep{Ravichandran2014,Ravichandran2017}.  
In the limit $St \to 0$ however, some general results can be obtained for all $N$. Indeed,  
if the fluid flow has a stagnation point at some position $\mathbf X_{eq}^0$, and  if the gradient of  
$\mathbf u$ at $\mathbf X_{eq}^0$ is invertible, the implicit functions theorem implies that  there exists a continuously differentiable function $\mathbf X_{eq}(St)$, defined for sufficiently small $St$, and satisfying   $\mathbf \Phi(\mathbf X_{eq}(St),St)=\mathbf 0$,  and $\mathbf X_{eq}(0) = \mathbf X_{eq}^0$. Particles with a sufficiently small Stokes number will then have an equilibrium position in the vicinity of the fluid stagnation point. Let $\mu$ denote an eigenvalue of the fluid stagnation point. The characteristic polynomial of the fluid gradient matrix at $\mathbf X_{eq}^0$ reads (see Appendix \ref{appI0}):
\begin{equation}
\mu^2 = \left(\frac{\partial u}{\partial x} \right)^2 + \left(\frac{\partial u}{\partial y} \right)^2 - 2  \frac{\partial u}{\partial y}   \equiv I(\mathbf X_{eq}^0)
\label{eqmu}
\end{equation}
where we made use of incompressibility and of  irrotationality of the flow in the lab frame.

To study the nature of the equilibrium position of inertial particles, we re-write their motion equation    (\ref{eqmvt}) as a dynamical system with 4 degrees-of-freedom for $\mathbf Y = (x,y,\dot x,\dot y)^T$. The system takes the form $\dot{\mathbf Y} = \mathbf F(\mathbf Y)$.
Let $\lambda_1,...,\lambda_4$ denote the eigenvalues of $\nabla\mathbf  F$ at $\mathbf X_{eq}(St)$ (this matrix is given in Appendix \ref{appI0}). These eigenvalues are
\begin{eqnarray}
    \lambda_{1,3} &=&\frac{1}{2 St}\left( -1+ (1-4 St^2 \pm 4 St \sqrt{I})^{1/2}\right),\\
      \lambda_{2,4} &=& - \frac{1}{2 St}\left( 1+ (1-4 St^2 \pm 4 St \sqrt{I})^{1/2}\right),
\end{eqnarray}
where $I=I(\mathbf X_{eq}(St))$. On expanding $I$ in the limit of small Stokes numbers, i.e. $I=I_0 + St I_1 + St^2 I_2 + ...$, with ${I_0}=I(\mathbf X_{eq}^0)$, we get:
\begin{eqnarray}
\label{lam1general}
    \lambda_{1,3} &=& \pm\sqrt{I_0} +\left(-1-I_0 \pm\frac{I_1}{2\sqrt{I_0}}\right) St + O(St^2)\\
      \lambda_{2,4} &=& - \frac{1}{St} \mp \sqrt{I_0}  + \left(1+I_0 \mp\frac{I_1}{2\sqrt{I_0}}\right) St + O(St^2) .
      \label{lam2general}
\end{eqnarray}
These eigenvalues give crucial information about the behavior of particles in the vicinity of the fluid equilibrium position. If $I_0 >0$, then the fluid stagnation point is a hyperbolic saddle (with eigenvalues $\mu = \pm \sqrt I_0$), and Eqs.\ (\ref{lam1general})-(\ref{lam2general}) show that the particle equilibrium point is also a hyperbolic point with four real non-zero eigenvalues, two of them being positive. It is therefore unstable.

If $I_0 < 0$, then the fluid   stagnation point   is an elliptic centre point (with eigenvalues   $\mu = \pm i\sqrt {|I_0|}$). Eqs.\ (\ref{lam1general})-(\ref{lam2general}) show that the particle equilibrium point is still hyperbolic, i.e. with complex eigenvalues with  non-zero real parts. These real parts can be readily calculated and read
\begin{eqnarray}
\label{lam1I0neg}
    Re(\lambda_{1}) &=&  Re(\lambda_{3}) = -\left(1+\mu^2 \right) St + O(St^2)\\
      Re(\lambda_{2}) &=& Re(\lambda_{4}) =  - \frac{1}{St} + \left(1+\mu^2\right) St + O(St^2) .
      \label{lam2I0neg}
\end{eqnarray}
Eqs.\ (\ref{lam1I0neg})-(\ref{lam2I0neg}) show that, if $\mu^2 < 0$ {\it and}  $\mu^2 > -1$, the equilibrium point $\mathbf X_{eq}(St)$ is asymptotically stable and attracts particles released in its basin of attraction, in the limit $St \to 0$. 
 
The determination of elliptic fluid stagnation points and the calculation of $\mu^2$ is therefore a quick way to detect possible trapping points of inertial particles in the flow. This is the method adopted in the present work and presented in the following paragraphs for particles in vortex crystals.

\subsection{Anticyclonic cells and trapping in the vortex crystal}

\begin{figure}
\centerline{\includegraphics[width=0.85\textwidth]{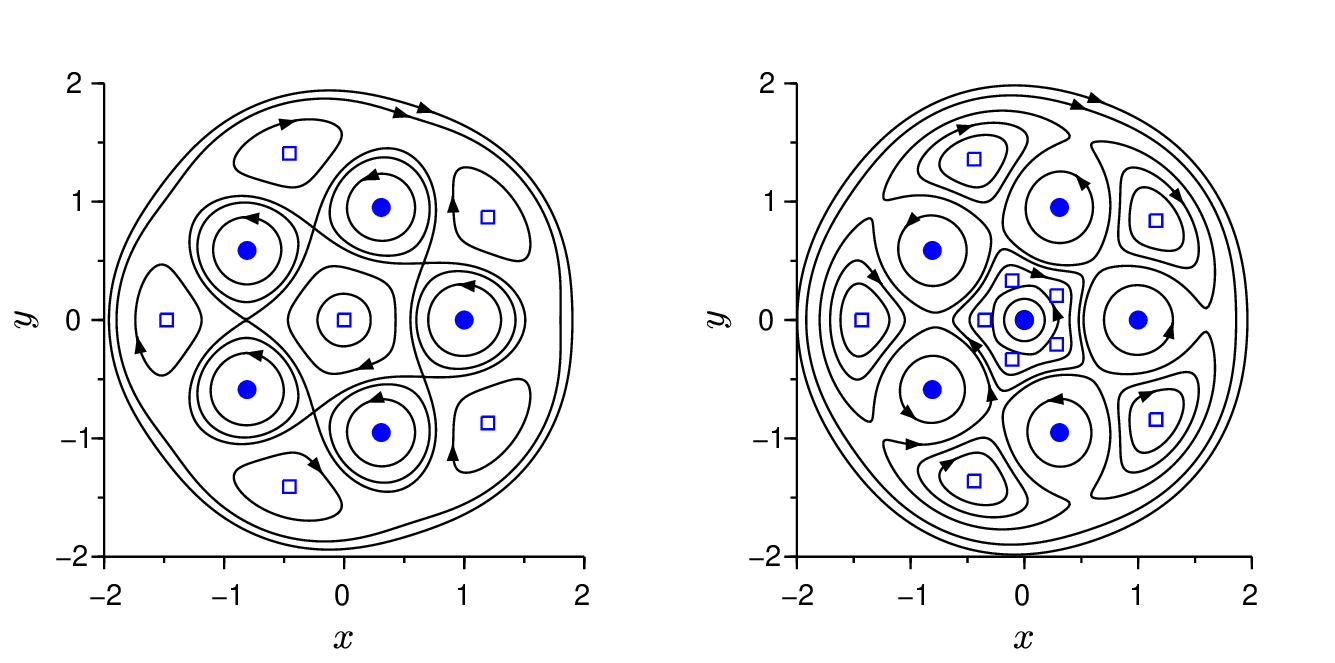}}
\caption{Relative streamlines of the flow created by $N=5$ vortices with identical strengths $\Gamma$ placed on a pentagon. Blue disks indicate the position of  vortices, squares are elliptic stagnation points at the center of anticyclonic cells. Left: no central vortex, $N+1= 6$ anticyclonic cells are present. Right: a vortex with strength $\Gamma_c = \Gamma/4$ has been added at the center, $2 \, N = 10$ anticyclonic cells are present. }
\label{LdC_seules_N5+1}    
\end{figure}

\begin{figure}
\centerline{\includegraphics[width=0.8\textwidth]{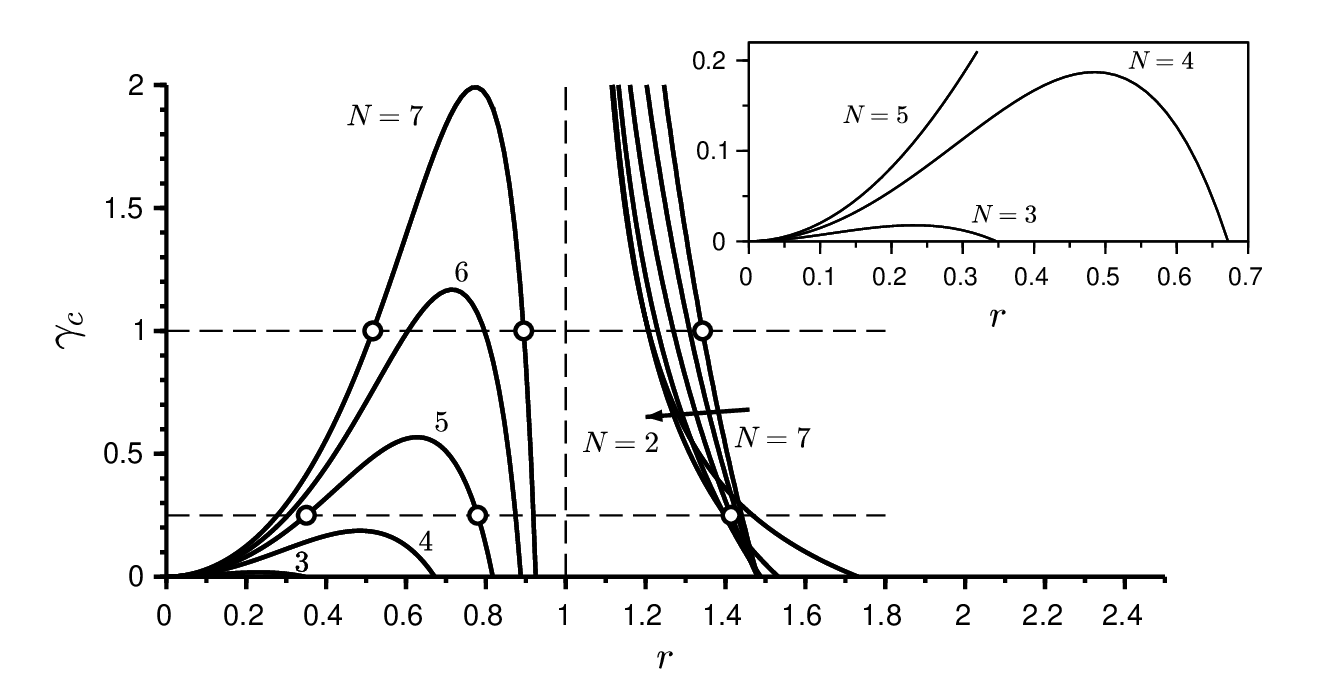}}
\caption{Lines $P_N(r,\gamma_c)=0$, for various values of $N$, indicating the existence of fluid stagnation points along the axis $\theta=\pi/N$. 
The inset is a zoom of the same graph.
The axis $\gamma_c=0$ corresponds to the case without central vortex, where $N+1$ elliptic stagnation points exist (one at $r=0$, plus $N$ external points at $r > 1$). When $\gamma_c > 0$, Eq.\ $P_N(r,\gamma_c)=0$ can have 3 solutions $r > 0$ for $3 \le N \le 7$, corresponding to $3N$ stagnation points, $2N$ of which are elliptic points. When $\gamma_c$ is above some critical value,   internal points vanish.  }
\label{PN_r_gammac}       
\end{figure}

The inspection of the relative streamlines $\psi = constant$ in the case $\gamma_c = 0$ shows that the relative flow has a large anticyclonic cell around an elliptic   stagnation point located along the axis $\theta = \pi/N$ $[mod \, 2\pi/N]$ (see left graph of Fig.\ \ref{LdC_seules_N5+1} in the case $N=5$). In the case $\gamma_c > 0$, new stagnation points appear along the very same axes. Indeed, one can check that the velocity components $u=\partial \psi/\partial y$ and $v= -\partial \psi/\partial x$, calculated  along the axis $\theta = \pi/N$ read
\begin{eqnarray}
u &=& f_N(r) \, P_N(r,\gamma_c),\\
v &=& g_N(r) \, P_N(r,\gamma_c),
\end{eqnarray}
where $f_N(r)$ and $g_N(r)$ are non-zero functions and $P_N(\gamma_c,r)$ is a polynomial function of $r$ 
given in Appendix \ref{secapp1}. The system $(u,v)=(0,0)$  is therefore equivalent to
\begin{equation}
P_N(r,\gamma_c) = 0.
\label{PNzero}
\end{equation}
Figure \ref{PN_r_gammac} shows the lines $P_N = 0$ in the plane $(r,\gamma_c)$, for various values of $N$. We observe that, for $\gamma_c = 0$ and $3 \le N \le 7$, only two stagnation points exist along the axis $\theta = \pi/N$, one at $r=0$ and one at $ r > 0$. Therefore, $N+1$ stagnation points exist in the $(x,y)$ plane (one at the center, plus $N$ external points at $r > 1$).  The  stagnation point  at $r=0$ is  a saddle point when $N=2$ and an elliptic point at the center of an anticyclonic cell when $3 \le N \le 7$ \citep{Angilella2024PoF}. 
When $\gamma_c > 0$ and $N \ge 3$, Eq.\ (\ref{PNzero}) can have up to 3 solutions $r > 0$, corresponding to $3N$ stagnation points in the $(x,y)$ plane. The 3 roots of $P_5(r,1/4)$ and 
$P_7(r,1)$ 
are marked with a white circle in Fig.\ \ref{PN_r_gammac}, and are denoted $r_1 < r_2 < r_3$ in the following. Using a numerical solver to determine $r_i$ for $N=5$ and $7$, we can determine $\mu^2$ at the fluid stagnation points $(r_i \cos(\pi/N),r_i \sin(\pi/N))$. The result is given in Table \ref{tablemu}.
 \begin{table} 
 \begin{center}
\begin{tabular}{ccccc} 
   & & ~~$\mu^2(r_1)$~~& ~~  $\mu^2(r_2)$~~ &~~ $\mu^2(r_3)$   \\ 
   \hline
  $N=5$\\ $\gamma_c=1/4$& & -0.72 & 3.72 & -0.92   \\
  \hline 
  $N=7$\\ $\gamma_c=1$ & & -0.69 & 4.29 & -0.90 \\  
    \hline
\end{tabular} 
 \caption{Squared eigenvalues of the fluid stagnation points $(r_i,\theta=\pi/N)$ identified on Fig.\ \ref{PN_r_gammac} (circles), for $(N,\gamma_c) = (5,1/4)$ and $(N,\gamma_c) = (7,1)$ .}
 \label{tablemu}
 \end{center}
\end{table}
We therefore conclude that attracting points exist for $(N,\gamma_c) = (5,1/4)$ and $(N,\gamma_c) = (7,1)$ in the vicinity of
$(r=r_1,\theta=\pi/N)$ and $(r=r_3,\theta=\pi/N)$, for inertial particles
with a sufficiently small Stokes number. In contrast, the point $(r=r_2,\theta=\pi/N)$
is a saddle stagnation  point in both cases, and no particle trapping point is expected to  exist in its vicinity under the current hypotheses.
 Since the same conclusions hold for $\theta=n\pi/N$, $n=1 .. N-1$, we conclude that $2N$ particle trapping points exist in these flows. Half of these $2N$ points are located inside the crystal ($r < 1$) and half of them are located outside ($r > 1$). 
We also note from Fig.\ \ref{PN_r_gammac} that, when $\gamma_c$ is above some critical value that can be readily determined (see table \ref{tablegamma_max}), the internal points vanish. In contrast, the external stagnation points are robust and persist for all $\gamma_c$. 
 \begin{table} 
 \begin{center}
\begin{tabular}{cccccccc} 
 $N$  & & 2 &  3 & 4 & 5 & 6 & 7   \\ 
   \hline
   $\gamma_c^{max} $& &  $\times$ &  0.0178 & 0.187 & 0.568 & 1.168 & 1.992    \\
    \hline
\end{tabular} 
 \caption{Critical value of the central vortex strength ratio $\gamma_c$ below which internal anticyclonic cells exist.}
 \label{tablegamma_max}
 \end{center}
\end{table}

\section{Transport and capture of clouds of particles}
\label{secnum}
The analysis presented in the previous section shows that particles must be trapped in anticyclonic cells, provided their Stokes number is sufficiently small. To verify this, we have solved numerically the transport equation of $10^4$ inertial particles in the laboratory frame, together with the inviscid transport equation of vortices under their mutual influence, in the cases $(N,\gamma_c) = (5,1/4)$ and $(N,\gamma_c) = (7,1)$. The left graph of Fig.\ \ref{Nuages_20toursN5+1} shows the cloud of particles (red points) after 20 turns of the crystal, together with the relative flow streamlines, in the former case. \tb{A movie of the evolution is available in the Supplementary Material.}
We observe that particles accumulate in 10 zones, in the vicinity of 5 external points ($r>1$) and 5 internal points ($r<1$), as expected. These correspond to the attracting points predicted by the linear analysis of the previous section. For longer times, one would see particles converging towards these points, and the red patches of particles would become much smaller. Similar conclusions hold for simulations in the case $(N,\gamma_c) = (7,1)$ (left graph of Fig.\ \ref{Nuages_20toursN7+1}).

Figures \ref{Nuages_20toursN5+1} and \ref{Nuages_20toursN7+1} also suggest that some particles converge towards a star-shaped line located in the vicinity of a streamline. Inertial particles at low Stokes numbers have been observed to accumulate near attracting streamlines in various plane flows. 
This point is discussed in the next section.

\section{Asymptotic identification of the attracting streamline}
\label{secattract_LdC}

A necessary condition for the determination of such attracting streamlines has been proposed by \cite{Haller2010}. It consists in setting to zero the rate-of-change of 
the volume enclosed by the attractor, leading to
\begin{equation}
      \int \dot\mathbf{X} \cdot \mathbf{n} \, ds = 0
\end{equation}
where the integral is taken over the attracting line,  $ds$ denotes the elementary arc-length, and $\mathbf{n}$ is a unit  vector perpendicular to the attracting line. This condition is particularly useful if one makes use of the classical perturbation of Eq.\ (\ref{eqmvt}) in the limit of small inertia
\begin{equation}
    \dot\mathbf{X} = \mathbf{u}   + St \left( -\nabla \mathbf u \cdot \mathbf u + \mathbf X - 2 \mathbf e_z \times \mathbf u \right).
\end{equation}
In addition, we assume that the attracting streamline is of equation $\psi(x,y) = \psi(x_0,0)$, i.e. it crosses the $x$ axis at some position $(x_0,0)$. By making use of the fact that $ds = | u | dt = | \nabla \psi | dt$ 
and $\mathbf n =   \nabla \psi /| \nabla \psi| $,
the criterion can be recast as a temporal integral
\begin{equation}
    J(x_0) \equiv \int_0^{T(x_0)} \left( -\nabla\psi \cdot \nabla \mathbf u \cdot \mathbf u + \nabla\psi \cdot\mathbf X - 2 |\nabla\psi|^2 \right) dt = 0,
\end{equation}
  where $T(x_0)$ is the period of fluid points along the streamline $\psi  = \psi(x_0,0)$. Note that $J(x_0)$ also corresponds to the difference of fluid streamfunction at two stroboscopic positions of an inertial particle $\mathbf X(0)$ and $\mathbf X(T(x_0))$. If $J(x_0)$ is non-zero the particle moves away from this streamline. 

\begin{figure}
\centerline{\includegraphics[width=\textwidth]{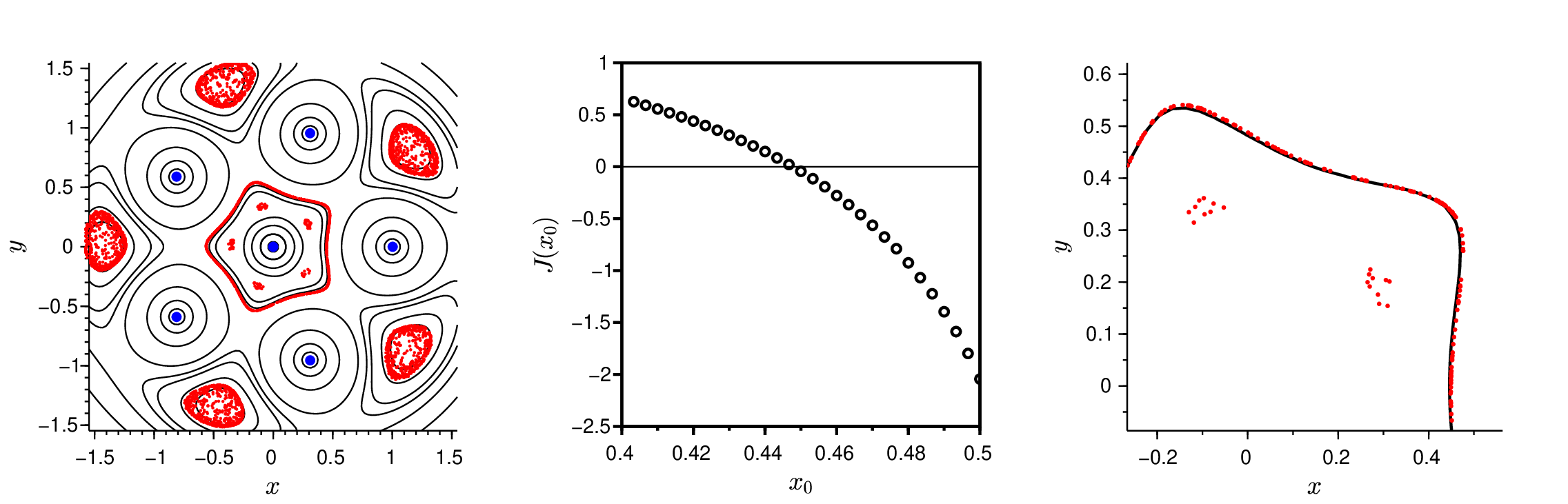}}
\caption{Left: global view of the cloud of particles (red dots) after 20 turns of the crystal (blue disks), together with fluid streamlines (solid lines), in the case of the inviscid computation of vortices forming a pentagone with a central vortex of relative strength $\gamma_c=1/4$. The accumulation of particles is visible near 5 external trapping points, 5 internal trapping points, and in the vicinity of an attracting streamline of equation $\psi(x,y)=\psi(x_0^*,0)$. Middle: plot of the streamfunction variation $J(x_0)$, such that $J(x_0^*)=0$ (Haller \& Sapsis criterion). The right graph shows the theoretical attracting streamline (black solid line), together with computed particles.  The Stokes number is $St=0.02$. \tb{See also the movie in the Supplementary Material.}}
\label{Nuages_20toursN5+1}       
\end{figure}

\begin{figure}
\centerline{\includegraphics[width=\textwidth]{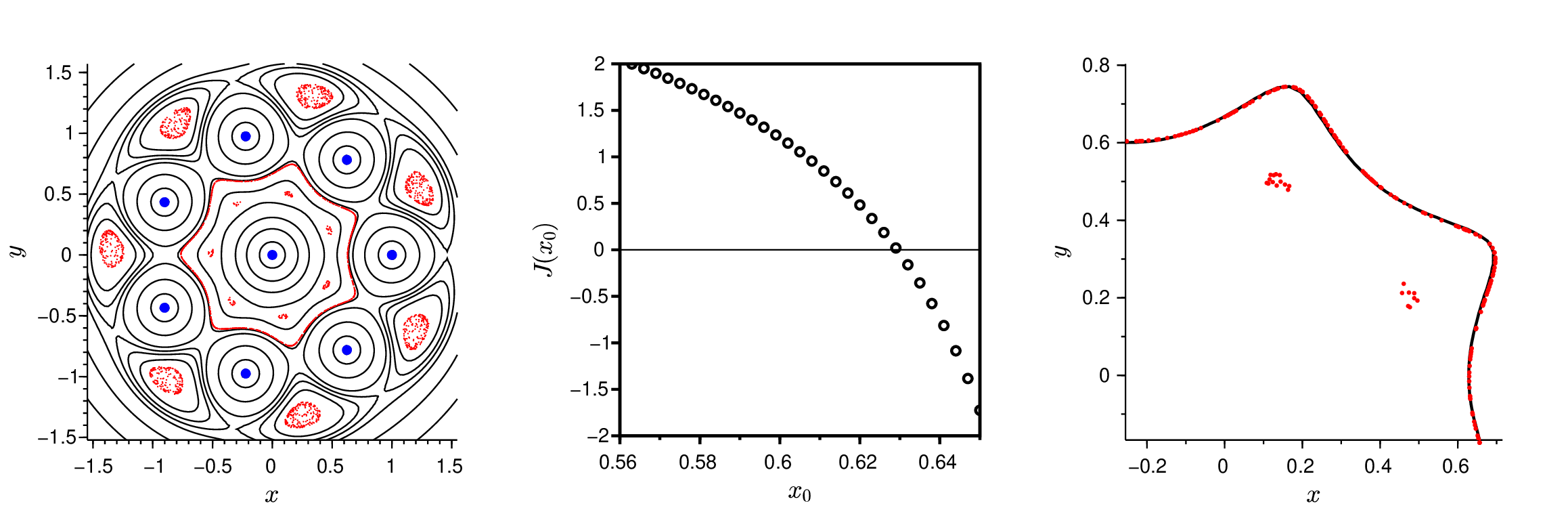}}
\caption{Counterpart of Fig.\ \ref{Nuages_20toursN5+1}, with $N=7$ and $\gamma_c=1$.}
\label{Nuages_20toursN7+1}       
\end{figure}

The central graphs of Figs. \ref{Nuages_20toursN5+1} and \ref{Nuages_20toursN7+1} show a computation of $J(x_0)$ in the cases $N=5$ and $7$. We observe that $J$ has a simple zero at $x_0^* \simeq 0.447$ ($N=5$) and  $x_0^* \simeq 0.629$ ($N=7$).  Particles indeed accumulate near the theoretical streamline  $\psi=\psi(x_0^*,0)$ (black solid line, right graphs of Figs. \ref{Nuages_20toursN5+1} and \ref{Nuages_20toursN7+1}).
At long times they form a mass crystal, in that the attractor is fixed in the rotating frame, but this crystal has dimension one, in contrast with the mass crystal formed in the absence of the central vortex. 

\begin{figure}
\centering
\includegraphics[width=1\columnwidth]{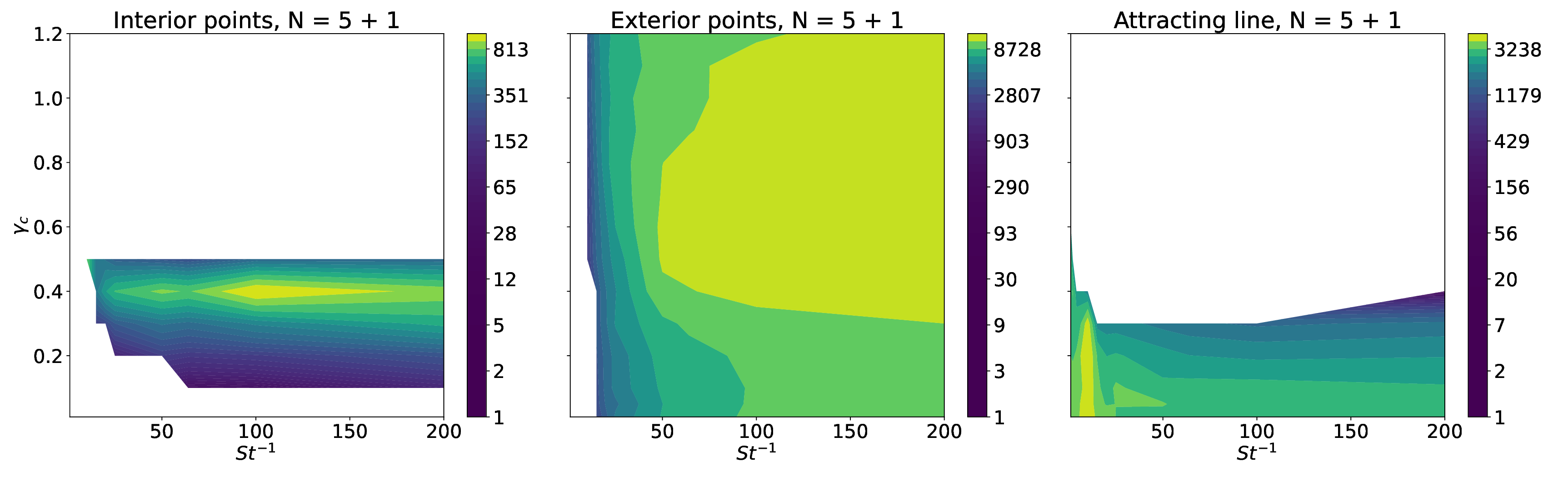}
\includegraphics[width=1\columnwidth]{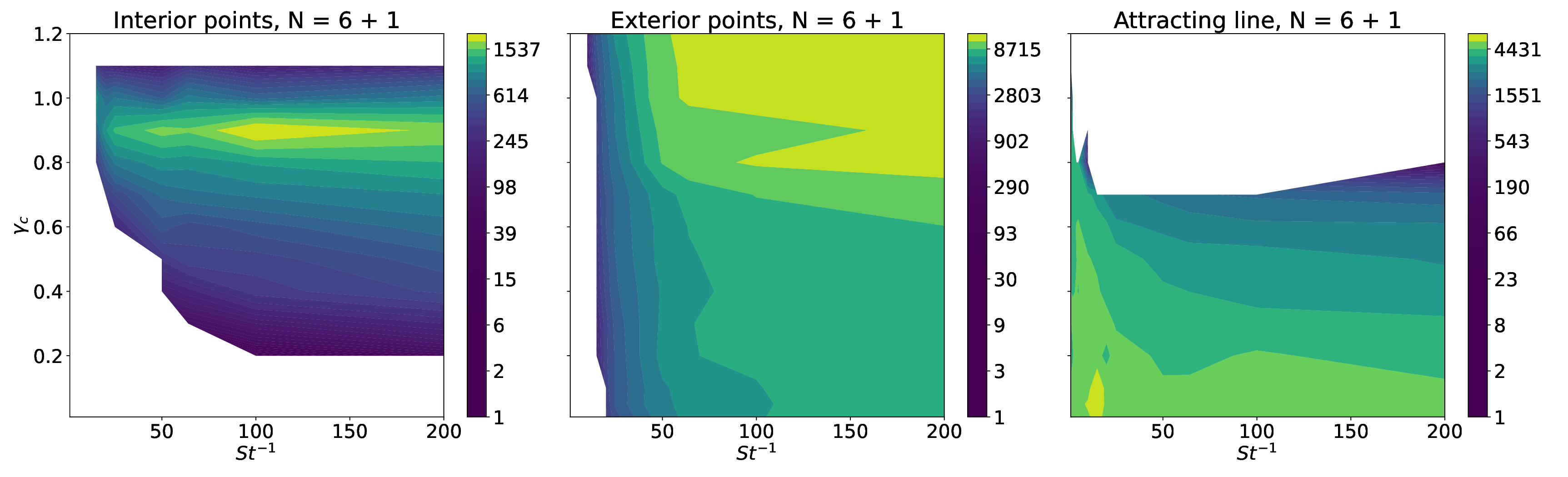}
\caption{\tb{\label{fig:counts} The numbers of particles trapped at (left) interior and (middle) exterior fixed points, and (right) the attracting streamline in simulations with (top) $N=5+1$ vortices and (bottom) $N=6+1$ vortices. Parameter combinations for which no particles are trapped are coloured white. The critical $\gamma_c$ above which the interior fixed points disappear may be read off the figures on the left and compared with Table \ref{tablegamma_max}. The total number of particles is $5\times10^4$ and $6\times10^4$ in the two cases (initialised with uniform number density in the region $r\leq3$), the colorbar is logarithmic, and particle counts smaller than $10$ are discarded. }}
\end{figure}

\tb{\subsection{Particle trapping by vortex crystals} \label{sec:particle_trapping}}
\tb{In general, the critical values of $\gamma_c>0$ for which the attracting streamline disappears and for which the interior or exterior fixed points disappear are different. In Fig. \ref{fig:counts}, we show that the combination of $\gamma_c$ and $St$ determines whether the interior or exterior fixed points of the attracting streamline can trap particles. The irregular boundaries of the regions of parameter-space in which indefinite trapping can occur belie the complex flow field in the vicinity of a vortex-crystal. }
We note that the attractor can be multi-pronged in the case where $\gamma_c < 0$ (i.e. when the central vortex has circulation of the opposite sense to the other vortices). However, at the magnitudes $|\gamma_c|$ for which the attractor is seen to split into two or more attractors, the vortex crystal in unstable in viscous simulations, and we do not discuss this case further.

\section{Effect of viscosity} \label{sec:viscous}

The potential flows used in the previous sections correspond to an idealized inviscid fluid where vorticity is localized into regions of null volume. Also, the corresponding velocity fields are steady (in the rotating frame) and persist for infinite times. These two conditions are no longer true when viscosity is present: under the effect of viscous diffusion, vortex cores will grow and vortex merger will eventually occur. However, all previous analyses concerning vortex pairs have shown that attractors exist temporarily in the presence of viscosity, provided the flow Reynolds number is large \citep{Angilella2010, Ravichandran2014, Angilella2014}.   Here, using direct numerical simulations (DNS) of the flow with Lagrangian particle tracking (the numerical method is detailed in Appendix \ref{sec:numerical_method}), we first show that the inviscid theory is robustly reproduced in viscous simulations,  in quantitative agreement with the inviscid case.  
Simulations were performed with $1024^2$ gridpoints and a  timestep of $\delta t = 0.005$ \tb{(nondimensionalised as in \S 2).}

In addition, we observe that for $\gamma_c = 0$, different long-term evolutions emerge and have important consequences on the particle concentration field. These various behaviours, which depend on the flow Reynolds number,   are described in the next sections.

\subsection{Crystals with a central vortex}
\begin{figure}
\centering
\includegraphics[width=1\linewidth]{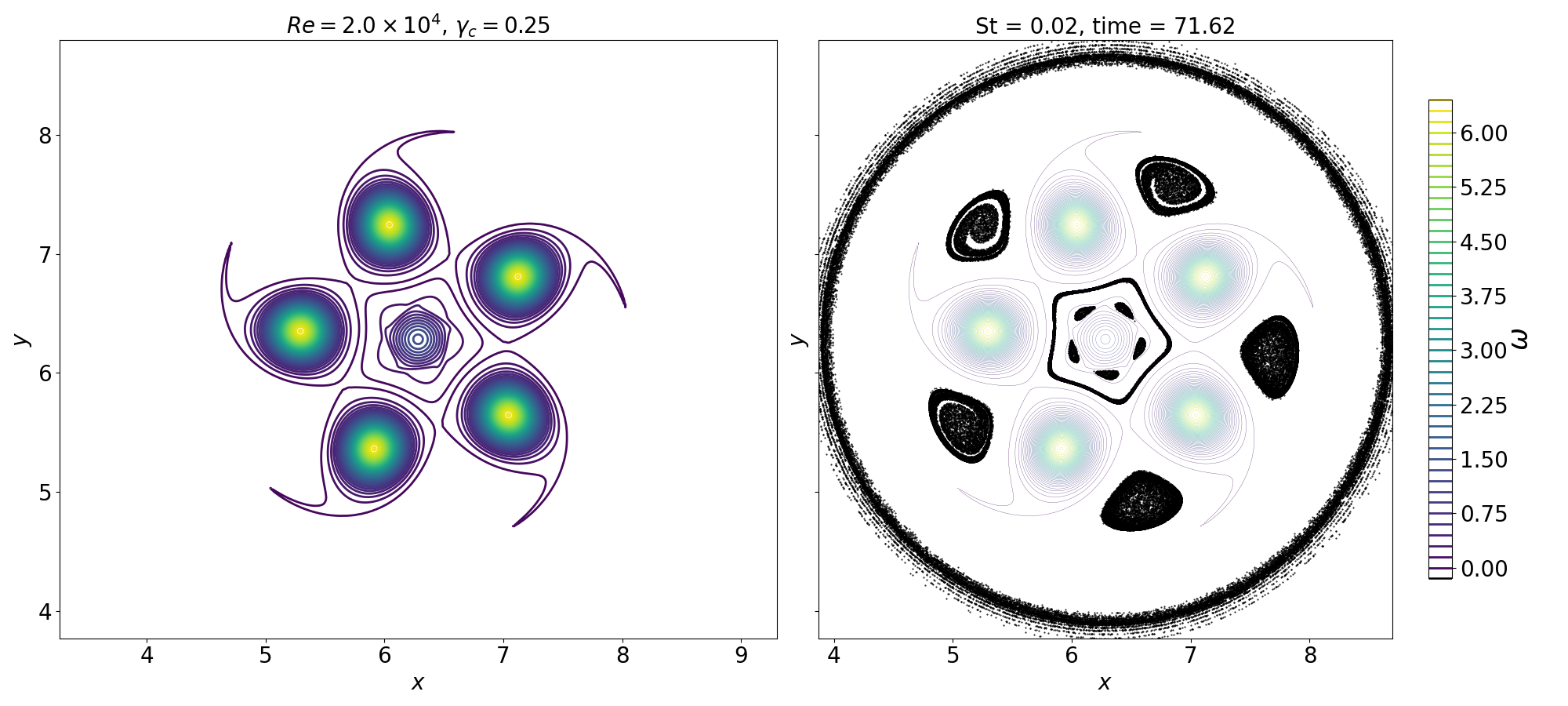}
\includegraphics[width=1\linewidth]{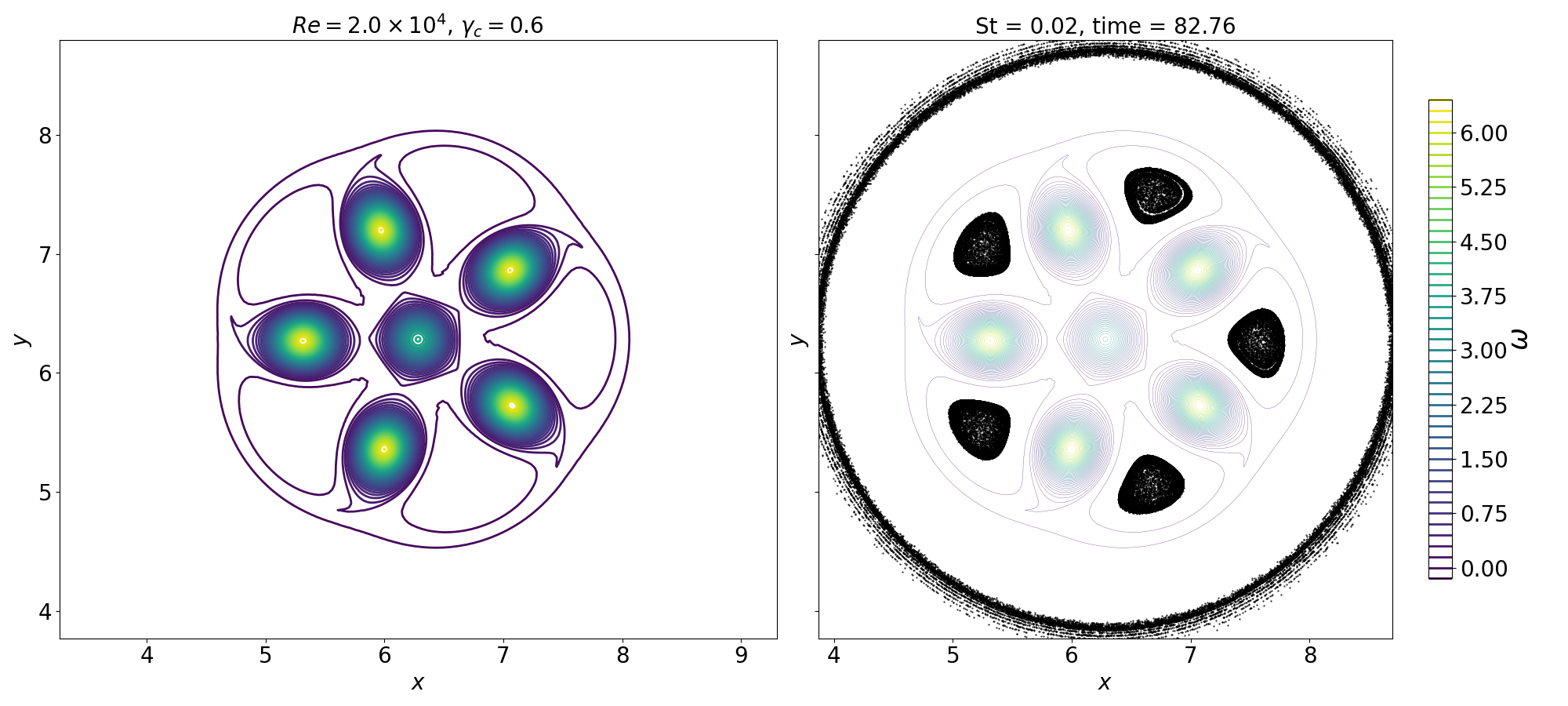}
\caption{Iso-contours of vorticity (left) and particle locations overlaid with iso-contours of vorticity (right).
Top: the interior and exterior fixed points as well as the attracting streamline are seen with $N=5$ vortices of strength $\Gamma = 1$, with a central vortex of strength $\gamma_c = 0.25$ for particles of Stokes number $St = 0.018 \approx 0.02$. The fixed points also exist for $St=0.036$ (not shown). Bottom: For $\gamma_c = 0.6 > \gamma_c^{max}$ (see Table \ref{tablegamma_max}), only the exterior fixed points are seen for any $St$. The Reynolds number $Re=\Gamma / \nu = 2\times 10^4$ (see Appendix \ref{sec:numerical_method} for details). Compare Figs. \ref{LdC_seules_N5+1} and \ref{Nuages_20toursN5+1}, \tb{together with the movie in the Supplementary Material.}}
\label{fig:5vor}
\end{figure}

Figure \ref{fig:5vor} shows contours of the vorticity and particle positions for simulations initialised with $5+1$ vortices of strengths $(\Gamma, \gamma_c)$. The interior fixed points exist and the crystal of particles is seen for $\gamma_c < \gamma_c^{max} = 0.568$, as expected (see Table \ref{tablegamma_max}). \tb{A movie of the evolution is available in the Supplementary Material.} 

\begin{figure}
    \centering
    \includegraphics[width=1\linewidth]{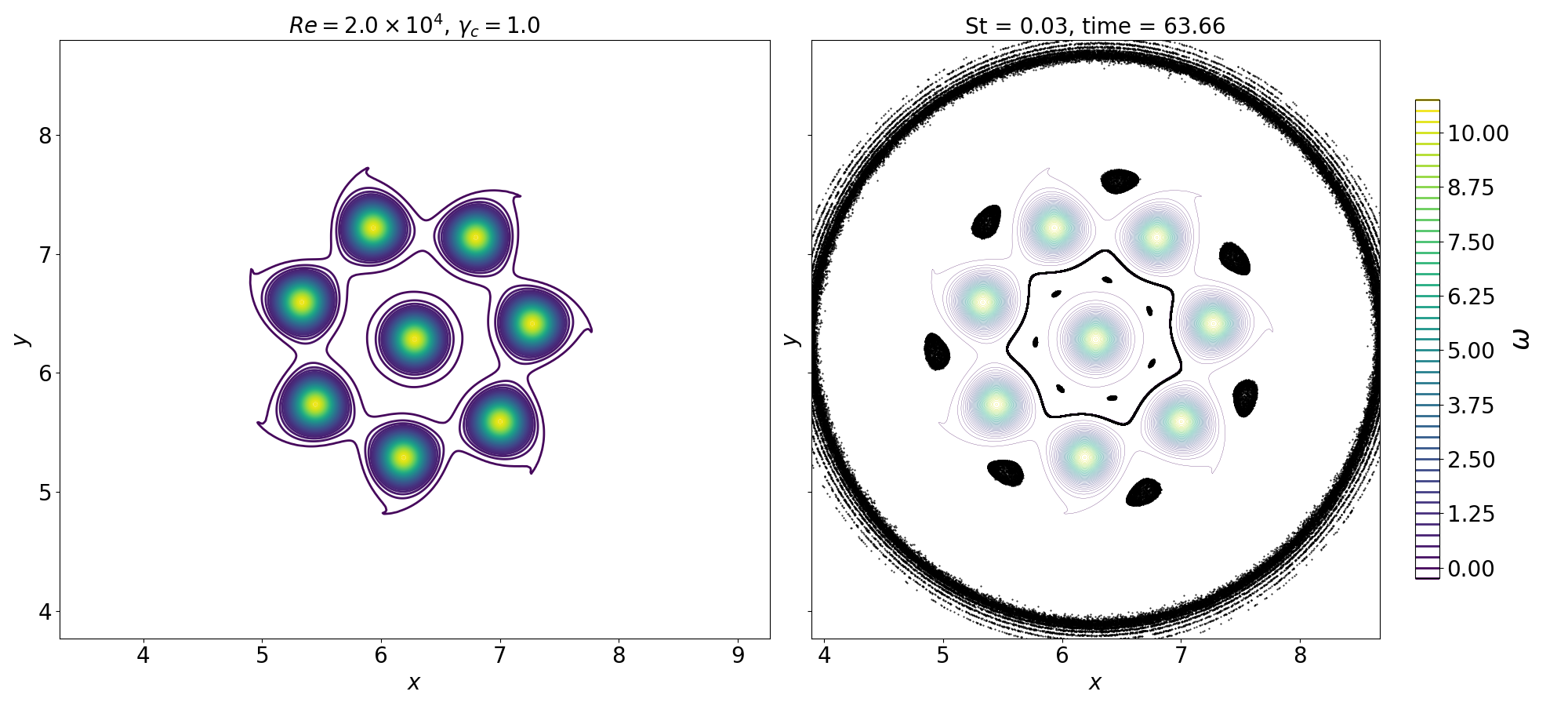}
    \includegraphics[width=1\linewidth]{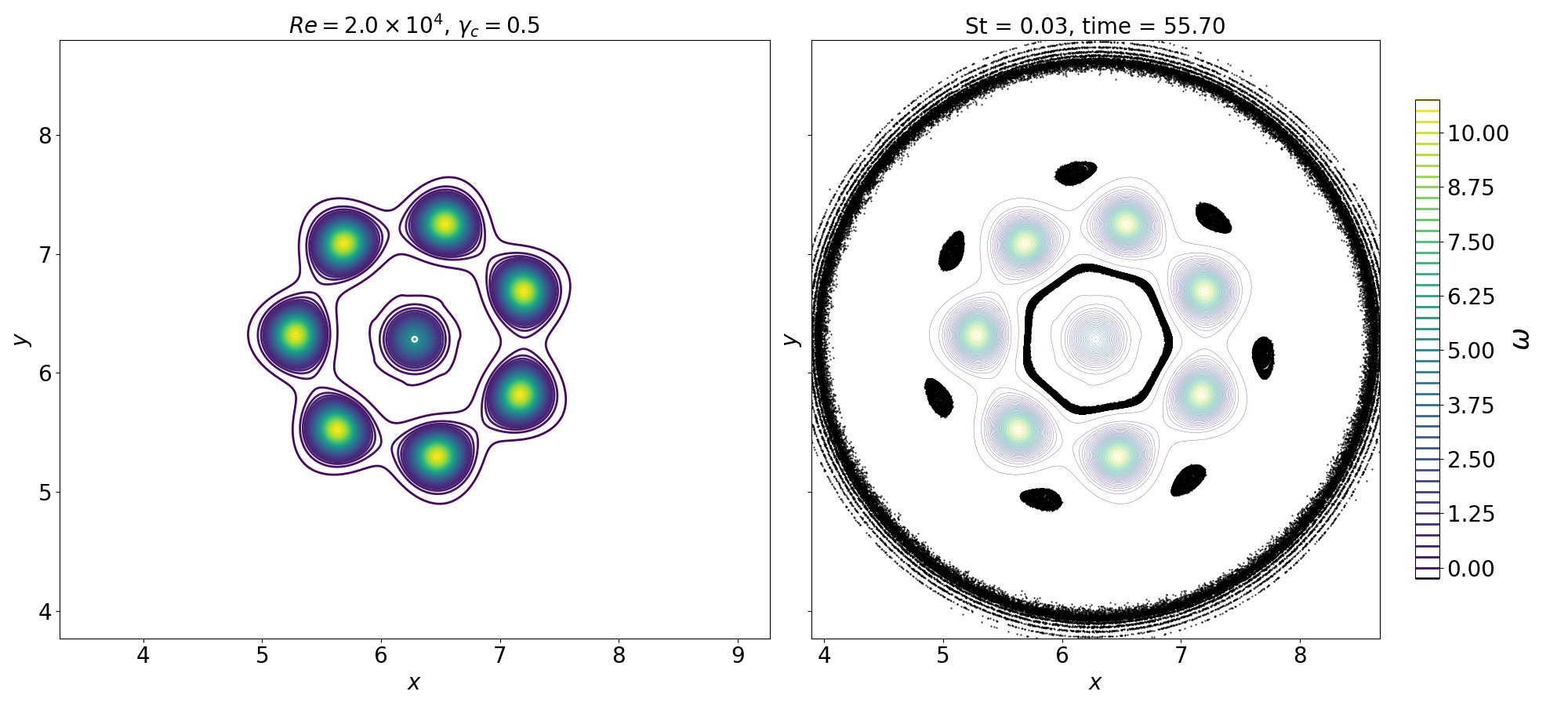}
    \caption{Same as Fig. \ref{fig:5vor}, but with $N=7$ vortices with (top) the central vortex strength $\gamma_c=1.0$ and Stokes number $St=0.032$
    \tb{(see also the movie in the Supplementary Material)}, and (bottom) $\gamma_c=0.5$ and $St=0.028$. While the attracting streamline is seen in the latter, the interior fixed points are not.}
\label{fig:7vor}
\end{figure}

In Fig. \ref{fig:7vor}, we plot the analogues of Figs. \ref{fig:5vor} for $N=7$ vortices. As in Fig. \ref{Nuages_20toursN7+1}, the exterior and interior fixed points and the attracting streamline are seen for $\gamma_c = 1$. \tb{The time-evolution can be examined in the movie in the Supplementary Material.} However, with a smaller $\gamma_c=0.5$, we see only the attracting streamline and exterior fixed points. The Stokes number $St=0.028$ is evidently larger than the critical Stokes number $St_c$ for the internal fixed points. We have checked that the internal fixed points are seen briefly for $St=0.014$, and more prominently for $St=0.007$. 

\subsection{Crystals without central vortex  }

\subsubsection{Transition to a circular vortex layer}
\label{sec:moderate_Re}
\begin{figure}
\centering
\includegraphics[width=1\linewidth]{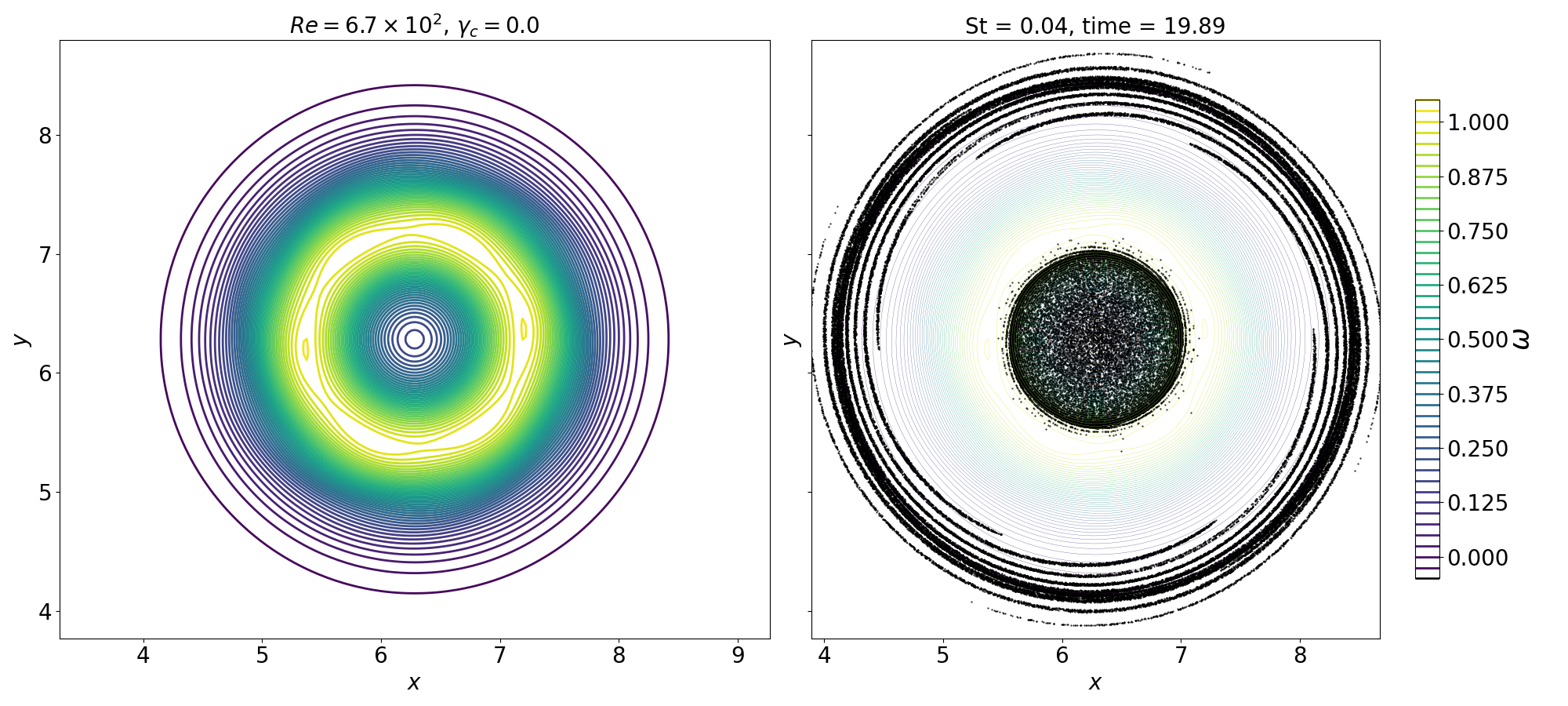}
\caption{Iso-contours of vorticity (left) and  particle locations overlaid with iso-contours of vorticity (right) after the merging of a $6$-vortex system, forming an annular vortex layer, at $Re=666$ without central vortex ($\gamma_c=0$). After merging, some particles remain inside the annular vortex, hence forming a dusty core. As very few particles escape from the dusty core, a  void zone expands at the exterior of the annular vortex layer where centrifugation takes place.}
\label{fig:6vor_annular}
\end{figure}

For Reynolds numbers $Re=O(10^3)$, systems of $N$ vortices (i.e. for $\gamma_c = 0$) are known to merge azimuthally to lead to an annular generalized Lamb-Oseen (GLO) vortex \citep[e.g.][]{swaminathan_2016}, which is a circular vortex layer with finite thickness. We observe that inertial particles suspended in such a flow  are ``corralled" by the annular vortex layer that is the result of the azimuthal merger of the vortices, as shown in Fig. \ref{fig:6vor_annular}. Since the diffusion of the annular vortex is a relatively slow process, particles remain trapped within the GLO for extended periods. The overall structure of the particle concentration field is close to the one obtained for particles released in an inviscid circular vortex sheet with zero-thickness. Indeed, for an axisymmetric  circular vortex sheet of radius $R$ and circulation $\Gamma_s$,   vorticity reads:
\begin{equation}
\omega(r) = \frac{\Gamma_s}{2\pi R} \, \delta(r - R),
\end{equation}
where $\delta$ is the Dirac distribution.
This flow corresponds to an azimuthal velocity $v_\theta(r) = 0$ for $r < R$ and  $v_\theta(r) = \Gamma_s/(2\pi r)$ for $r>R$. It can be thought of as a Rankine vortex where the solid core has been replaced by a still-fluid zone, thus creating a velocity discontinuity at $r=R$.  
If one releases particles in this flow with a uniform initial number density $n_0$ over a zone containing the whole disk of radius $R$, then the number density $n(r,t)$ at any time can be obtained by making use of the method of characteristics \citep[see][]{Druzhinin1994a}: for $r<R$ we have $n(r,t)=n_0$ for all $t$ ("dusty core" due to corralled particles), and for $R < r  < \rho(t) = R (1+\tau_p \Gamma_s^2 t/(\pi^2 R^4))^{1/4}$ a void zone, free of particles, develops ($n(r,t)=0$). The radius $\rho(t)$ corresponds to the position of particles released at $r \to R$, $r>R$, which are centrifuged away by the circular motion at the exterior of the circular sheet. Such particles are not replaced by particles coming from the the zone $r < R$, as no centrifugation exists in this zone,  in contrast with the Rankine vortex. 
\tb{In viscous flow, the vortex sheet evolves in the form \citep[e.g.][]{swaminathan_2016}
\begin{equation}
\omega(r,t) = \frac{\Gamma_s}{4\pi\nu t}e^{-(R^2+r^2)/4\nu t} I_0 \left(\frac{Rr}{2\nu t} \right)
\end{equation}
where $I_0$ is the modified Bessel function of the first kind and zeroth order.  }
Both the dusty core and the void zone are visible in our viscous simulations (Fig.\ \ref{fig:6vor_annular}).
For $r > \rho(t)$ a classical centrifugation takes place and concentration decays with $r$.  

 \begin{figure}
 \centering    
 \includegraphics[width=1\linewidth]{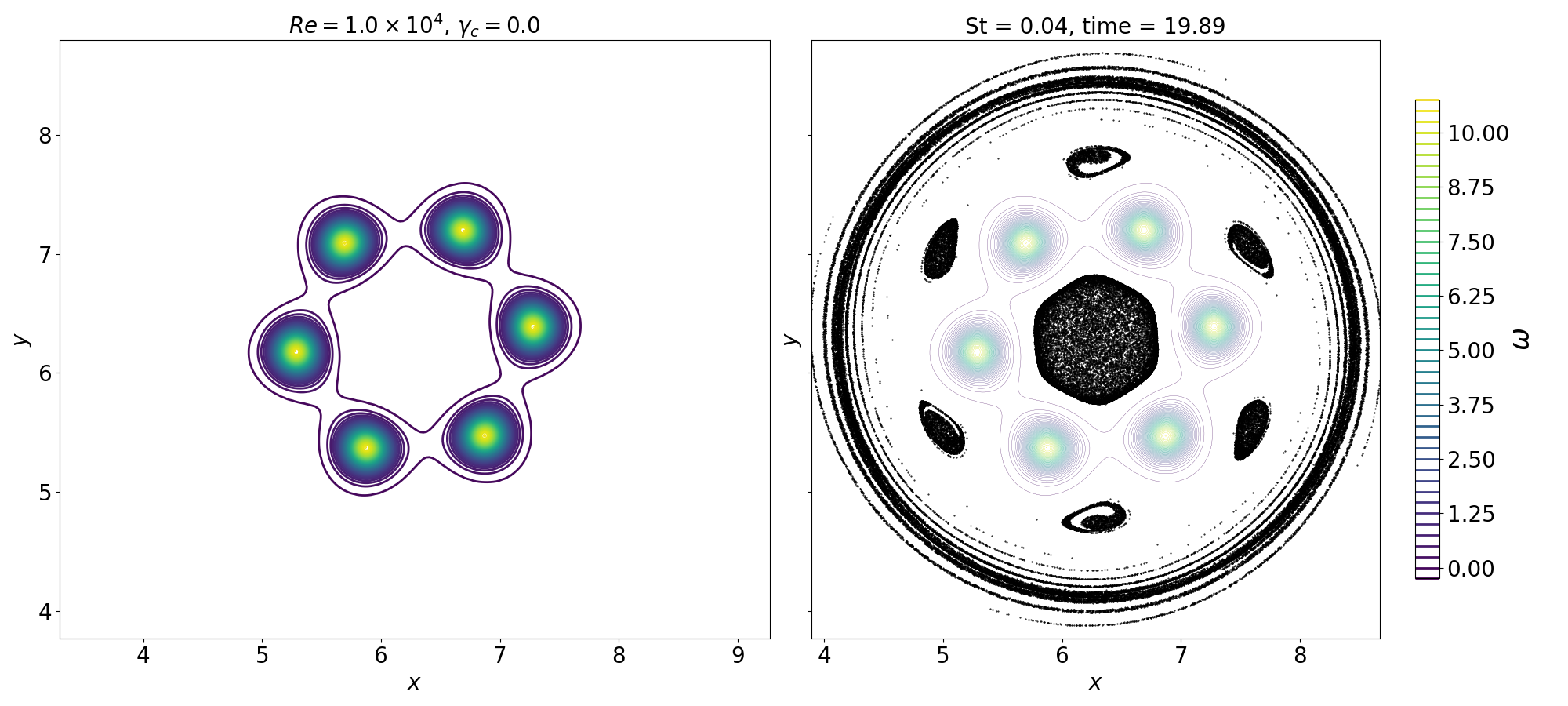}
 \caption{Iso-contours of vorticity (left) and  particle locations overlaid with iso-contours of vorticity (right), showing inertial particles of Stokes number $St=0.04$ trapped at the exterior fixed points and in the vicinity of the elliptic fixed point at the centroid of the $6$-vortex system with Reynolds number $Re=10^4$. The behaviour of the blob of particles near the origin is similar to that seen with an annular vortex (n.b. Fig. \ref{fig:6vor_annular}), but the distinct vortices lead to a tighter cluster at the origin. }    \label{fig:6vor_elliptic_trapping}
 \end{figure}
\begin{figure}
\centering
\includegraphics[width=0.49\linewidth]{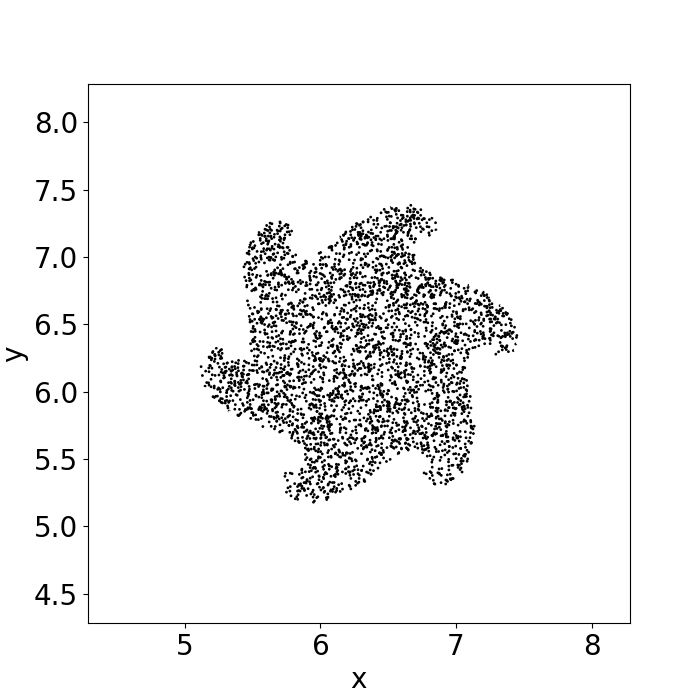}
\includegraphics[width=0.49\linewidth]{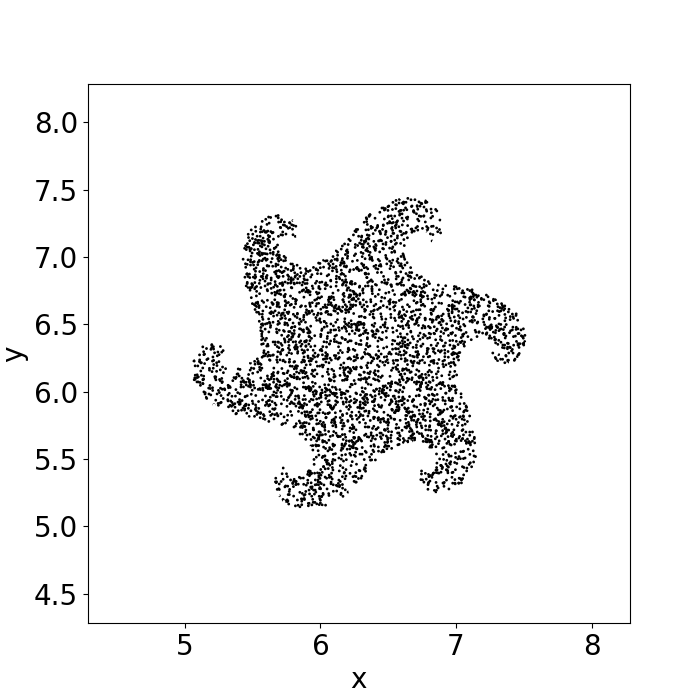}
\includegraphics[width=0.49\linewidth]{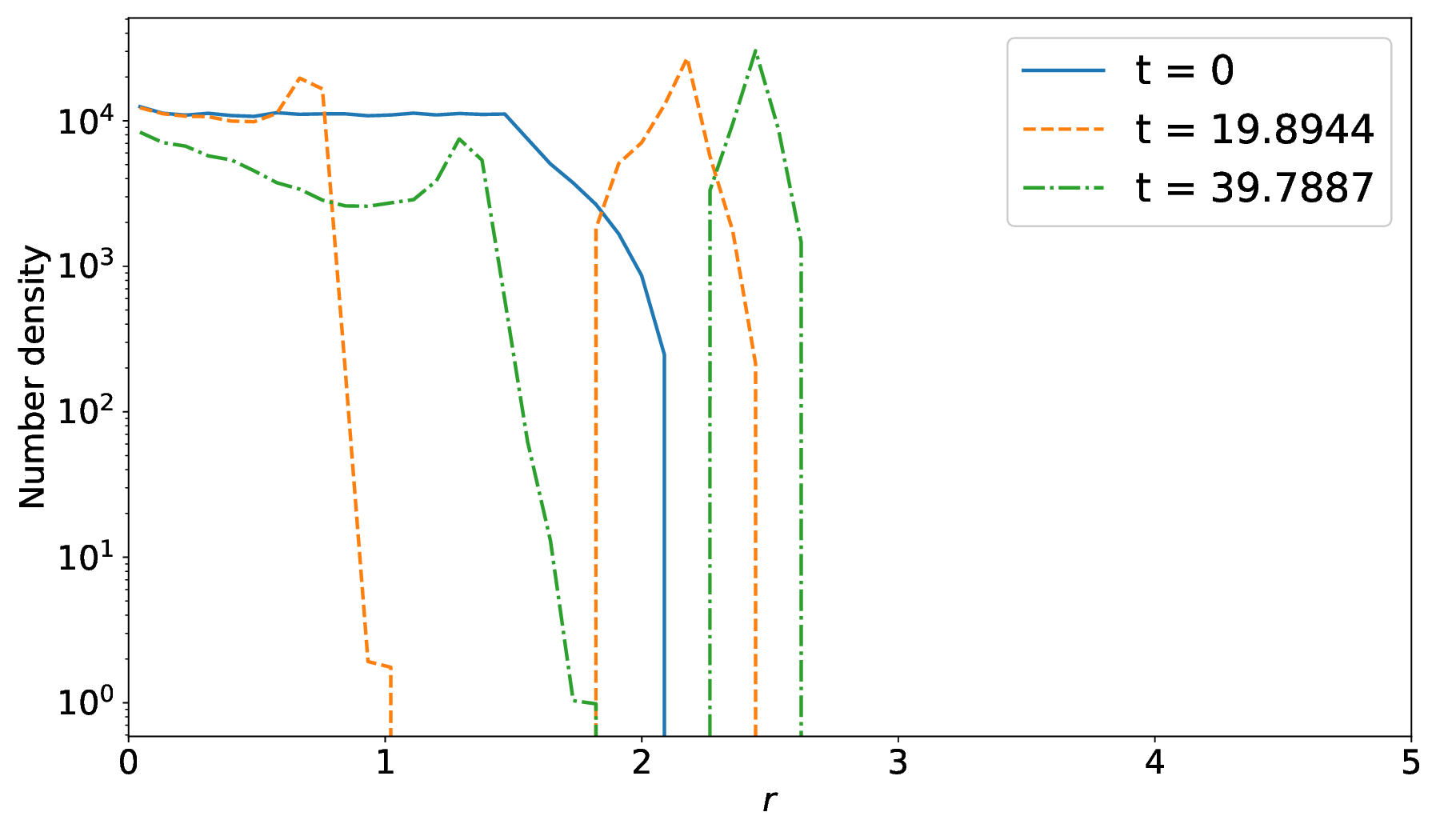}
\includegraphics[width=0.49\linewidth]{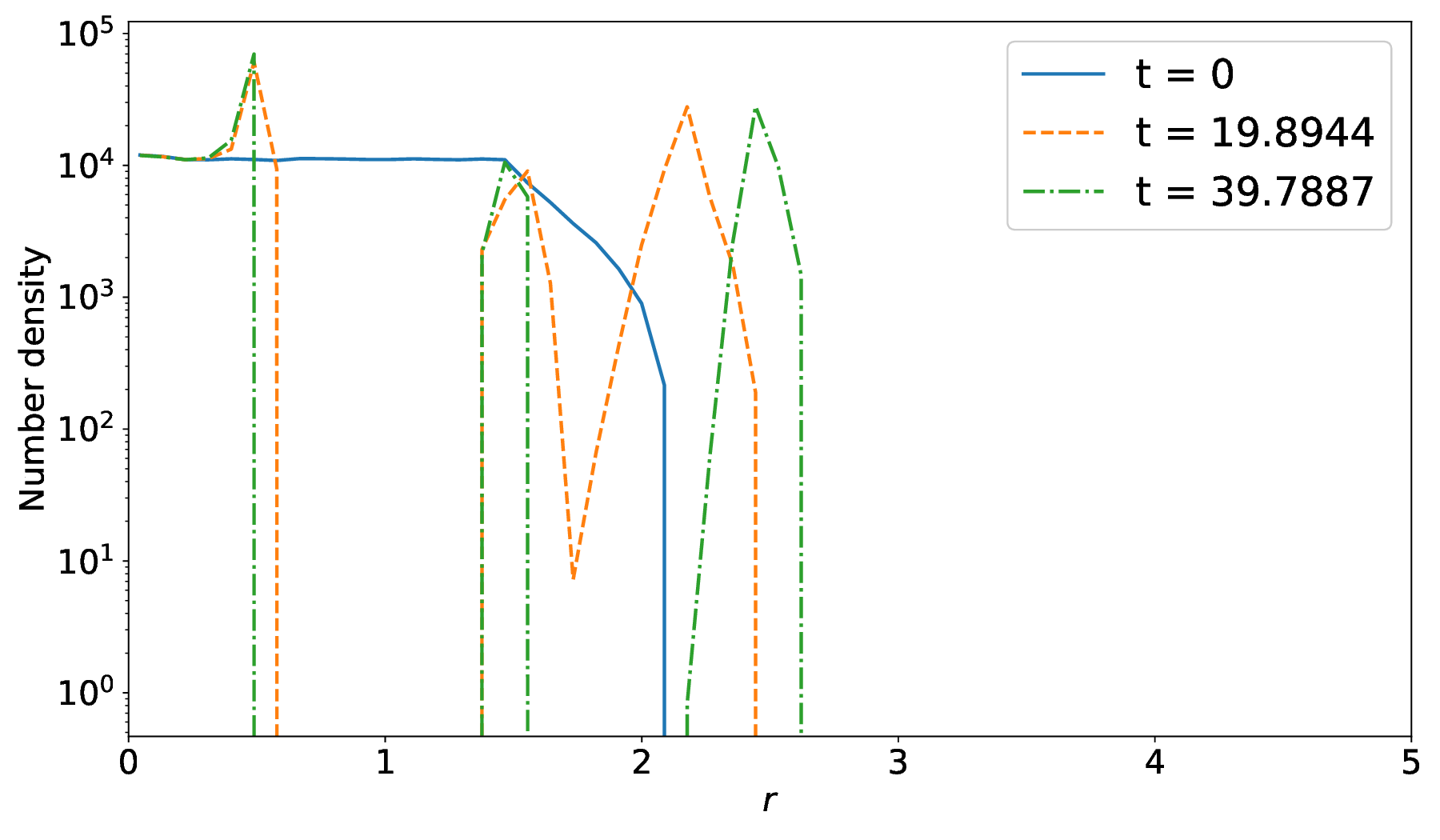}
\caption{Top: Basins of attraction of the particle blob at the origin, plotted at $t\approx20$, for the $6$-vortex system at $Re=666$ (left) and $Re=10^4$ (right). Bottom: Particle number densities for the $6$-vortex systems from Fig. \ref{fig:6vor_annular} at $Re=666$ (left) and $Re=10^4$ (right). The blob is more closely packed in the latter.}
\label{fig:6vor_par_dens}
\end{figure}

\subsubsection{Breakup  into a smaller crystal and persistence of trapping } \label{sec:breakup}

\begin{figure}
\centering
\includegraphics[width=1\linewidth]{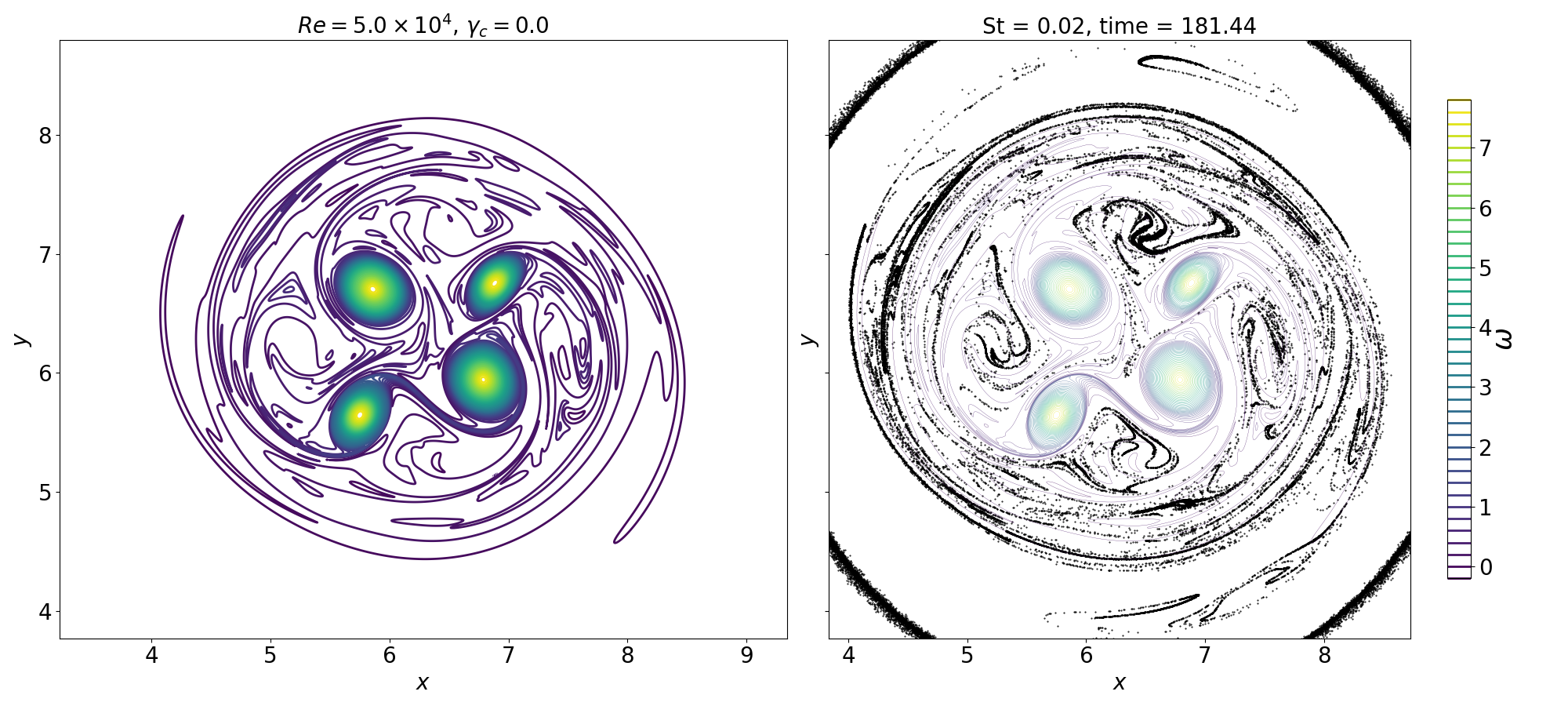}
\caption{Iso-contours of vorticity (left) and  particle locations overlaid with iso-contours of   vorticity (right), showing the intermediate $4$-vortex stage which results from the disintegration of a $7$-vortex system with $\gamma_c=0$, and persists for $\approx 30$ flow time units. Inertial particles are trapped at the (external) fixed points for the $4$-vortex system. The Reynolds number is $Re=5\times10^4$. Note that the vortices are of unequal strength in the intermediate stage.}
\label{fig:7vor_secondary_trapping}
\end{figure}

\begin{figure}
\centering
\includegraphics[width=0.8\linewidth]{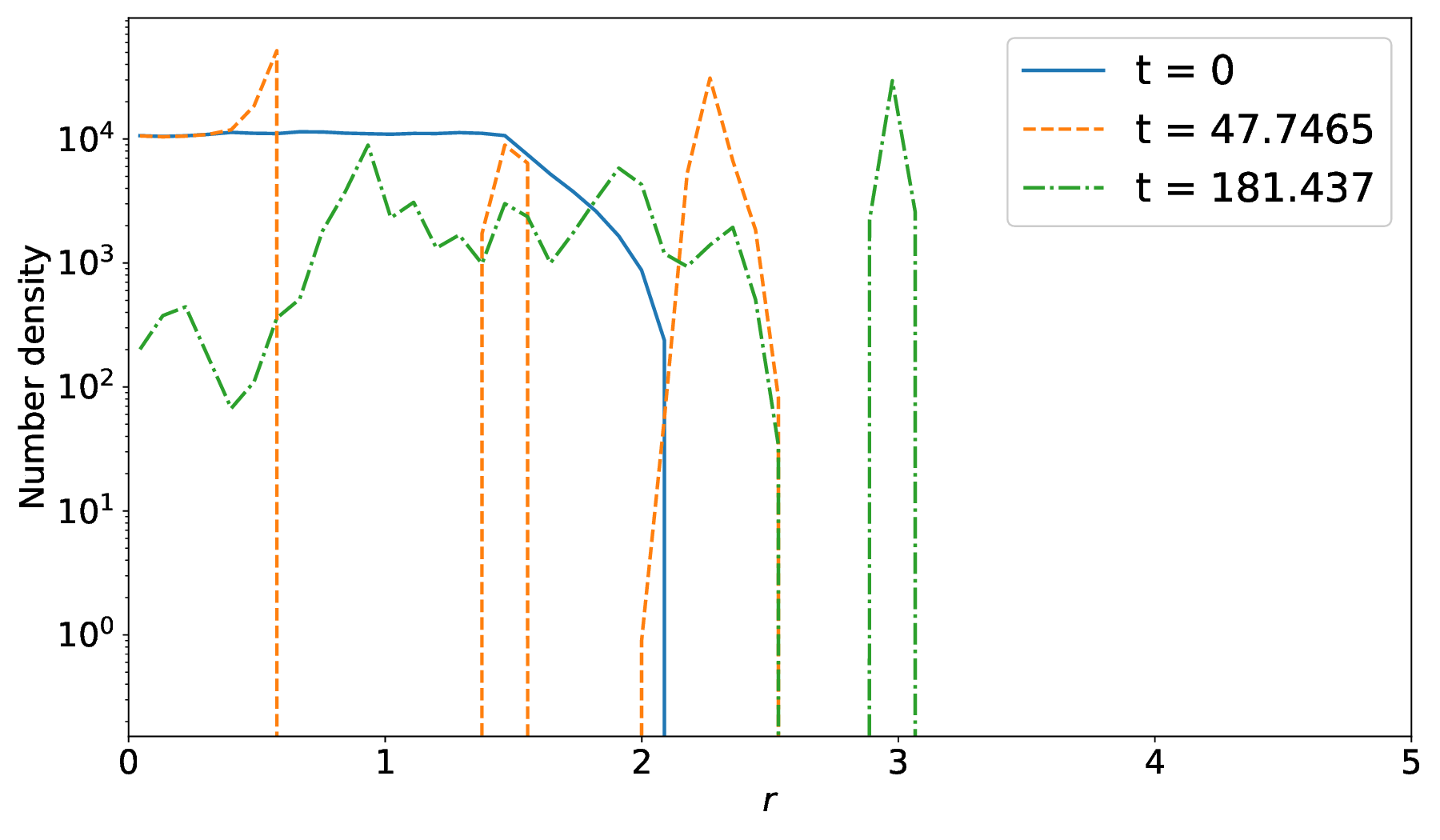}
\caption{The number density of particles per unit area as a function of the radius $r$ from the centroid of the vortex crystal for the simulation in Fig. \ref{fig:7vor_secondary_trapping}. At $t=0$, the number density is uniform for $0<r<1.5$ by construction. At $t\approx48$, we see three spikes. The peak at $r<1$ corresponds to increased particle density at the edge of the central blob (see also Fig. \ref{fig:6vor_annular}), and the peak at $r\approx1.5$ corresponding to the exterior fixed points. The outer ring of particles being centrifuged away are denoted by the peak at $r>2$. At $t\approx 181$, this outermost peak has moved to $r\approx3$; the peak at  $r\approx1$ corresponds to the particles at fixed points of the secondary $4$-vortex system. }
\label{fig:numdens_7vor_secondary}
\end{figure}

The $N$-vortex system ultimately evolves into a single vortex of positive sign at the origin. For the large Reynolds numbers considered here, the disintegration of the $N$-vortex crystal can, proceeding with pairwise mergers of vortices, lead to the formation of intermediate regimes with $M<N$ vortices. We observe that such $M$ vortex systems can also trap particles at the attracting fixed points associated with their geometry. In Fig. \ref{fig:7vor_secondary_trapping},
we show an example of this behaviour where the $7$-vortex crystal with $\gamma_c = 0$ and $Re=5\times10^4$ disintegrates and forms a $4$-vortex intermediate stage. This $4$-vortex system captures some of the particles from the elliptical fixed point at the centroid of the $7$-vortex system that are being centrifuged away, and particles are trapped at the fixed points of the (unequal) $4$-vortex system for a few dozen flow units. This is seen in the plot of the number density of particles in Fig. \ref{fig:numdens_7vor_secondary}, where the initially uniform areal distribution of particle number density is `unmixed' and particles aggregate at the interior and exterior fixed points ($t=48$). Once the $7$ vortices have merged into $4$ vortices ($t=181)$, particle trapping at the (external) fixed points of the $4$-vortex system is seen as a peak in the number density. 
Eventually, the smallest of the vortices in the $4$-vortex system is merged into one of the other vortices, quickly followed by the merger of all remaining vortices into a solitary vortex, and the trapped particles are centrifuged out to infinity.

Analogous behaviour is also seen in $N+1$-vortex systems (with $\gamma_c >0$), but fewer particles are trapped at the fixed points of the secondary $M$-vortex system since the blob of particles from where these particles are sourced is absent.

\section{Discussion and Conclusion} \label{sec:discussion}

We have shown that inertial particles released in an inviscid plane flow characterized by vortices forming a crystal have a variety of attracting positions. In the reference frame rotating with the crystal, these attracting positions are fixed points located in the vicinity of fluid's elliptic stagnation points, in the limit where the Stokes number is small. In the presence of a central vortex, an attracting streamline can also appear between the central and the satellite vortices, and the position of this streamline could be found by making use of asymptotic techniques \citep{Haller2010}. 

In the presence of viscosity, these behaviors have been shown to persist until vortex merger takes place. In systems of $N+1$ vortices, that is $N$ annular vortices plus one central vortex,  the particles are locally centrifuged away from the vortex at the centroid, but can still be trapped in the vicinity of annular vortices, and form an annular structure topologically similar to the attracting streamline seen in, e.g. Figs. \ref{fig:5vor} and \ref{fig:7vor}.

 After merging, the original crystal no longer exists, $\Omega_0$ is no longer defined, and different behaviors can happen.  
In the absence of a central vortex and for large but moderate Reynolds numbers, merging creates a robust annular vortex layer with still fluid (in the lab frame) inside a well-defined ring, and a decaying potential vortex flow outside this ring \citep{swaminathan_2016}. Particles located within the ring are ``corralled": they are kept inside this dead zone and form a dusty core for a large amount of time. Particles located outside this ring are centrifuged away, so that a void zone forms at the exterior of the vortex layer. For larger Reynolds numbers, vortex crystals have been observed to break-up into a smaller crystal (e.g. from 7 vortices to a 4-vortex system, Fig.\ \ref{fig:7vor_secondary_trapping}). There, simulations showed that particles could be trapped towards the fixed points of the 4-vortex system.

This work therefore shows that trapping of dust particles in vortex crystals can take different forms and appears in various situations. It can persist even when the vortex crystal evolves under the effect of viscosity. Although vortex merger eventually creates flows with very different structures (annular vortex sheet or smaller crystal), trapping of particles has been shown to persist under different forms.

When particles converge towards accumulation points, one expects inter-particle distances to decrease and particle concentration to grow. The former effect is likely to induce particle-particle interactions, and the latter might alter the fluid flow. Both effects are neglected here since we assumed non-interacting particles and one-way coupling for all times.  
Also, in the presence of viscosity and particles at semi-dilute concentrations, it was shown that vortex merger could be affected by the dispersed phase \citep{Shuai2024}.   
The inclusion of these complex effects in our models is a topic of great interest and is left for future studies.

Our results may be applicable to  geophysical settings which are characterised by systems of strong vortices.  For example, size-dependent preferential clustering of inertial particles in flows dominated by vortices is of interest in hurricanes, and may modify calculations for collisional charge-separation \citep{blackElectrification1999}. \cite{Sapsis2009} note that for typical flow speeds seen in hurricanes, inertial effects become significant for particles $O(10)$cm in size (which are likely to be pieces of debris rather than hydrometeors). Attracting points and limit cycles for inertial particles may be explicitly accounted for in such analyses. 
Studies of idealised eyewalls of hurricanes \cite[][see also \cite{swaminathan_2016} and references therein]{schubertPolygonal1999} have shown that an instability of the annular vorticity distribution can lead to polygonal vortex formations of the kind studied in detail in previous analyses \citep{swaminathan_2016, Angilella2024PoF} and in the present work.
The distribution of hydrometeors (collectively referring to all forms of solid and liquid water particles) in such hurricanes determines the resulting precipitation. Furthermore, collisions between ice particles of different sizes  \citep{Dash2001} in such systems lead to the charge separation that is responsible for lightning \cite[see][]{blackElectrification1999}.  
In these cases, the mechanisms of (extreme) inertial clustering as the one described here may modify the relevant estimates of collisions and coalescence leading to precipitation and lightning. 
  In these realistic situations, there might exist velocity components in the vertical direction. Also, gravity can force particles to sink. Both mechanisms can affect the dynamics of particles, so that  the pointwise attractors observed here might either be destroyed by these three-dimensional effects, or take the form of more complex attractors (e.g. attracting lines in the direction perpendicular to the main two-dimensional flow). Also, the attracting streamline discussed in \S \ref{secattract_LdC} might either vanish or take the form of a multi-dimensional attracting manifold. The detailed analysis of these effects will be considered in the near future.

The accumulation of inertial particles in anticyclones is a well-known phenomenon in astrophysical or geophysical flows \citep[e.g.][]{Barge1995,Tanga1996,beron-veraDissipative2015}, where the Coriolis force is induced by the rotation of the planet. The results of the present study show that trapping of objects in anticyclonic zones also appears when the Coriolis force is due to the rotation of a local system of vortices and is more common than may be realized.

\appendix
 
\section{Eigenvalues of fluid and particle equilibrium positions}
\label{appI0}

If $\mu$ is an eigenvalue of the fluid gradient matrix we have
$$
(u_{,x} - \mu)(v_{,y}-\mu) - v_{,x} u_{,y} = 0,
$$
where the coma indicates the derivation. We make use of incompressibility, i.e. $u_{,x}+v_{,y}=0$. Also, the flow in the laboratory frame is irrotational, so the relative velocity field $(u,v)$ has a uniform vorticity equal to $-2$ (as times have been set non-dimensional by $1/\Omega$), so $v_{,x} - u_{,y}=-2$. This leads to $$(u_{,x} - \mu)(u_{,x} + \mu) + (u_{,y}-2) u_{,y} = 0,
$$
and to Eq.\ (\ref{eqmu}).

The particle dynamics is written as a dynamical system with four degrees of freedom $\mathbf {\dot{Y}} = \mathbf F(\mathbf {\dot{Y}})$, where $\mathbf {\dot{Y}}=(x,y,\dot x,\dot y)^T$. 
The gradient of $\mathbf F$ at some equilibrium position $\mathbf X_{eq}(St)$ reads:
\[ 
\nabla \mathbf F = \left( \begin{array}{cccc}
0 & 0 & 1 & 0 \\
0 & 0 & 0 & 1 \\
\frac{1}{St} u_{,x}+1 & \frac{1}{St} u_{,y} & -\frac{1}{St} & 2 \\
\frac{1}{St} v_{,x}  & \frac{1}{St} v_{,y} + 1 & -2 & -\frac{1}{St}  
\end{array} \right)
\] 
 the eigenvalues of which are given by Eqs. \ (\ref{lam1general})-(\ref{lam2general}).

\section{Fluid stagnation points}
\label{secapp1}
The system $(u,v)=0$ is equivalent to $P_N(r,\gamma_c)=0$, where
$$
P_2(\gamma_c,r) = {r}^{4}-3\,{r}^{2}+2\gamma_c\, ({r}^{4}-1)
$$
$$
P_3(\gamma_c,r) = {r}^{5}-3\,{r}^{3}+{r}^{2}+\gamma_c ({r}^{5} - {r}^{3}+ 
\,{r}^{2}-1)
$$
$$
P_4(\gamma_c,r) =  3\,{r}^{6}-8\,{r}^{4}+3\,{r}^{2}+2\gamma_c\,({r}^{6}- {r}^{4} + {r}^{2}-1)
$$
$$
P_5(\gamma_c,r) =  2\,{r}^{7}-5\,{r}^{5}+2\,{r}^{2}+ \gamma_c \,({r}^{7}-{r}^{5} +  {r}^{2}-1)
$$
$$
P_6(\gamma_c,r) =  5\,{r}^{8}-12\,{r}^{6}+5\,{r}^{2}+ 2\gamma_c\,({r}^{8}-{r}^{6} +{r}^{2}-1)
$$
The polynomial $P_7(\gamma_c,r)$ has a much longer expression and will not be given here.

\section{Pseudospectral solver for viscous simulations} \label{sec:numerical_method}
In the laboratory frame, the Navier-Stokes equations nondimensionalised using the length and time scales discussed in the text,
\begin{eqnarray}
\boldsymbol{\nabla \cdot u = 0} \label{eq:continuity}, \mathrm{and}\\
\frac{D\boldsymbol{u}}{Dt} = -\nabla p + \frac{1}{Re} \nabla^2 \boldsymbol{u}, \label{eq:mom}
\end{eqnarray}
where $Re = \Gamma / \nu $ is the Reynolds number, and $\nu$ is the kinematic viscosity, are integrated in two spatial dimensions $(x,y)$ and time $t$ using a pseudospectral solver previously used in the same context \citep[e.g.][]{ravichandran2017b,ravichandran_orientation_2023}. 
The equations are integrated in time using a third-order low-storage strong-stability-preserving Runge-Kutta (SSPRK) method \citep{gottlieb_strong_2001}, modified to use Crank-Nicolson timestepping for the viscous terms. The nonlinear terms are dealiased by setting Fourier coefficients with $|k| > 2k_{max}/3$ to zero. 

The motion equation of heavy particles submitted to a linear Stokes drag read
\begin{equation}
    \frac{d\boldsymbol{v}}{dt} = \frac{\boldsymbol{u-v}}{St} \label{eq:maxey-riley},
\end{equation}
which is effectively Eq. (\ref{eqmvt}) in an inertial frame. The trajectories of particles are found by integrating Eq. (\ref{eq:maxey-riley}) in time with the second order exponential time-differencing (ETD2) scheme of \cite{Cox_Mathews_2002}. The fluid velocity at the particle location is obtained using bilinear interpolation.

Simulations are initialised with $N$ Gaussian vortices of strength $\Gamma=1$ and radius $r_v = 0.1$ on the vertices of a regular polygon in a doubly periodic box in the region $(0,4\pi)\times(0,4\pi)$, with or without the central vortex of relative strength $\gamma_c$ at $(2\pi,2\pi)$. A total of $10^5$ particles are initialised with the local fluid velocity in a region of size $3\times 3$ centred at the origin.

\begin{figure}
    \centering
    \includegraphics[width=1.0\linewidth]{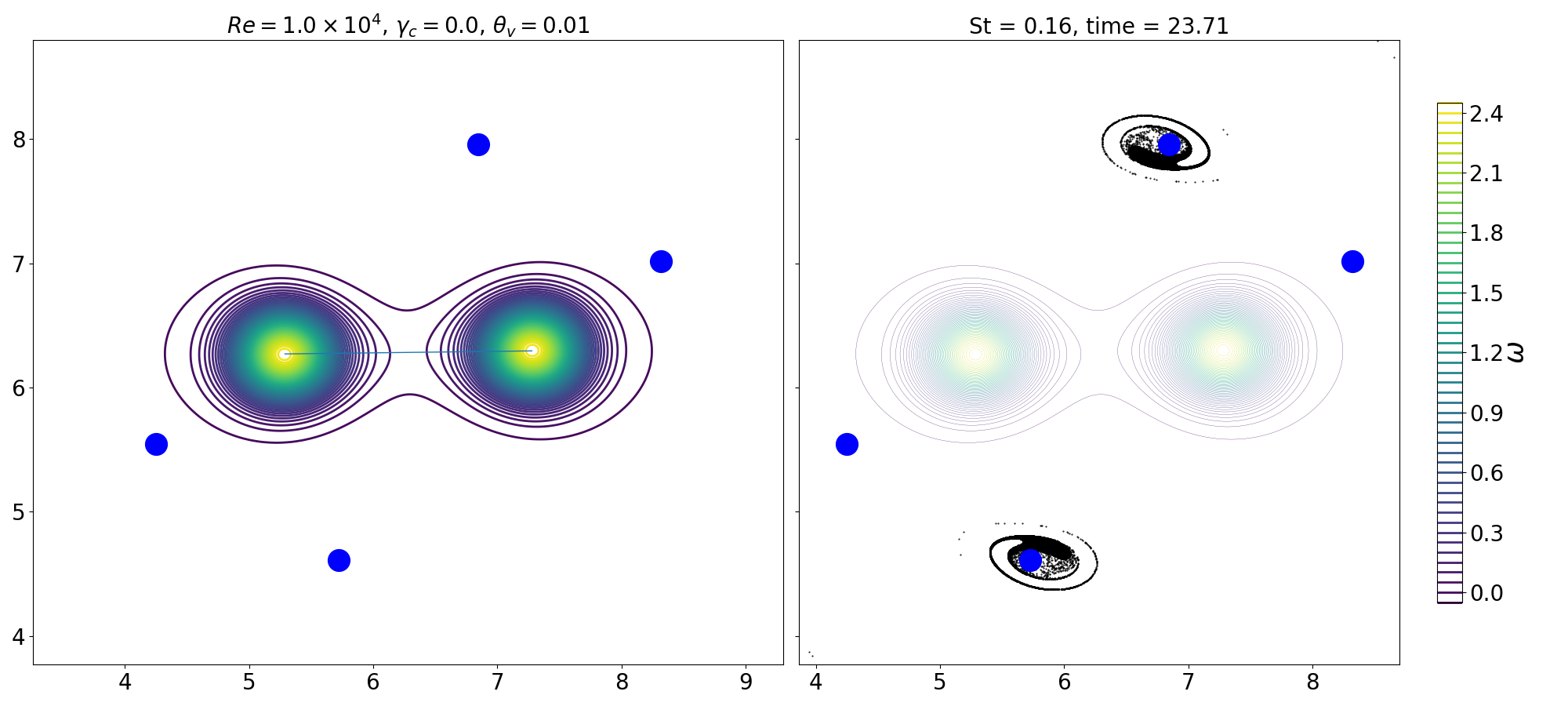}
    \includegraphics[width=1.0\linewidth]{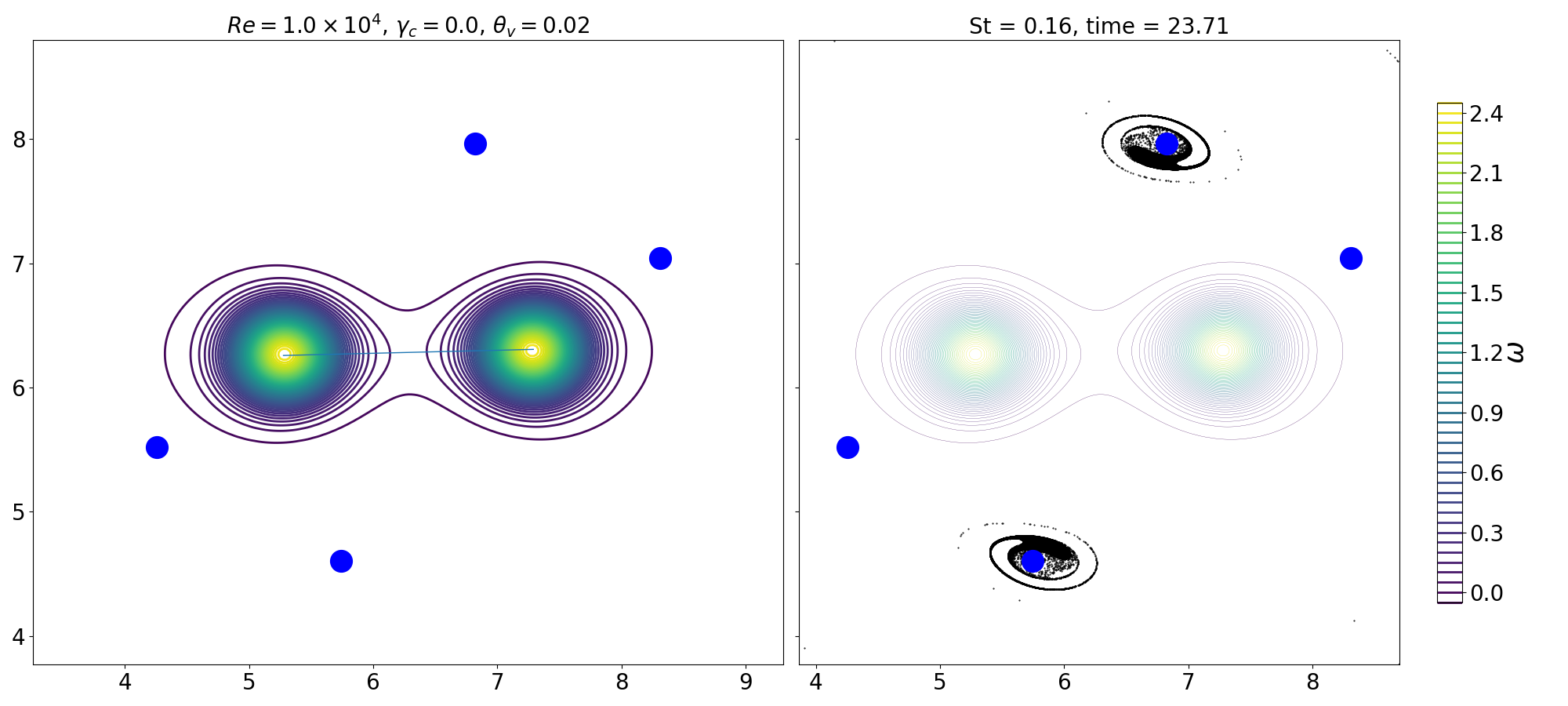}
    \caption{Particles of Stokes number $St=0.16$ (black dots) aggregate at the two stable fixed points around two vortices at a nondimensional time $t=23.71$ (in units of $\Omega_0^{-1}$; see text). The Reynolds number is $Re=10^4$. Top: results from a simulation with $1024^2$ grid points and timestep $\delta t=0.01$; Bottom: results from a simulation with $512^2$ grid points and $\delta t = 0.01$. The thin line in the figures on the left connects the centres of the vortices. The angle the thin line makes with the horizontal is displayed on top of each figure. The stable and unstable  fixed points from \cite{Angilella2010} are also shown (blue dots).}
    \label{fig:validation_2vor_tau2}
\end{figure}

The fluid solver has been previously validated in \cite[][see Supplementary Material]{ravichandran_orientation_2023}. Here we provide validation for the method used to integrate the particle equations of motion. In Fig. \ref{fig:validation_2vor_tau2}, we present a snapshot of a simulation with $N=2$ vortices, and $St=0.16$. For these parameters, attracting fixed points exist, as seen in \cite{Angilella2010} and \cite{Ravichandran2014}, and are recovered in the viscous simulations. 

\section*{Acknowledgements}
SR is supported through the IIT Bombay Seed Grant RD/0522-IRCCSH0-020, and through Core Research Grant no. ANRF/F/1930/2024-2025 from the Anusandhan National Research Foundation (ANRF).  Computations were performed on PARAM-RUDRA. We acknowledge National Supercomputing Mission (NSM) for providing computing resources of ‘PARAM RUDRA’ at G02, Ground Floor, IIT Bombay, Powai, Mumbai, 400076, which is implemented by C-DAC and supported by the Ministry of Electronics and Information Technology (MeitY) and Department of Science and Technology (DST), Government of India.

\section*{Declaration of Interests} The authors report no conflict of interest.


\end{document}